\documentclass[%
11pt,
 aip,
 jcp,%
 amsmath,amssymb,
reprint,%
]{revtex4-1}

\usepackage{pdfpages}
\usepackage{pgffor}
\usepackage{etoolbox} % fix rotated output with revtex4-1

\makeatletter
\patchcmd{\@outputpage@head}{\@ifx{\LS@rot\@undefined}{}{\LS@rot}}{}{}{}
\makeatother

\usepackage{graphicx}% Include figure files
\usepackage{dcolumn}% Align table columns on decimal point
\usepackage{bm}% bold math
\usepackage{pgfplots,listings,pgfplotstable,enumitem}
\usetikzlibrary{math,decorations.pathreplacing,calc}
\pgfplotsset{compat=newest}
\usepackage{stmaryrd}
\usepgfplotslibrary{fillbetween}

\pgfplotsset{
    legend entry/.initial=,
    every axis plot post/.code={%
        \pgfkeysgetvalue{/pgfplots/legend entry}\tempValue
        \ifx\tempValue\empty
            \pgfkeysalso{/pgfplots/forget plot}%
        \else
            \expandafter\addlegendentry\expandafter{\tempValue}%
        \fi
    },
}

\pgfplotstableset{fixed zerofill,precision=8}

    \usetikzlibrary{arrows, arrows.meta, 
                    chains,
                    positioning,
                    shapes}
\makeatletter
\tikzset{reset join/.code={\def\tikz@after@path{}}}
\makeatother

\usepackage{amssymb,amsmath,amsthm}

\usepackage{mathtools}

\usepackage{mathrsfs}
\usepackage{txfonts}
\usepackage{verbatim}
\usepackage{graphicx}
\usepackage{braket}
\usepackage{xr-hyper}
\usepackage{hyperref}

\makeatletter
\newcommand*{\addFileDependency}[1]{% argument=file name and extension
  \typeout{(#1)}
  \@addtofilelist{#1}
  \IfFileExists{#1}{}{\typeout{No file #1.}}
}
\makeatother

\definecolor{mygray}{rgb}{0.5, 0.5, 0.5}

\begin{document}

\preprint{}

\title{A state-specific multireference coupled-cluster method based on the bivariational principle}% Force line breaks with \\

\author{Tilmann Bodenstein}
\author{Simen Kvaal}%
 \email{simen.kvaal@kjemi.uio.no.}
\affiliation{Hylleraas Centre for Quantum Molecular Sciences, Department of Chemistry, University of Oslo, P.O. Box 1033 Blindern, N-0315 Oslo, Norway 
}%

\date{\today}% It is always \today, today,

\begin{abstract}
A state-specific multireference coupled-cluster method based on Arponen's bivariational principle is presented, the bivar-MRCC method. The method is based on singlereference theory, and therefore has a relatively straightforward formulation and modest computational complexity. The main difference from established methods is the bivariational formulation, in which independent parameterizations of the wavefunction (ket) and its complex conjugate (bra) are made. Importantly, this allows manifest multiplicative separability (exact in the extended bivar-MRECC version of the method, and approximate otherwise), while preserving polynomial scaling of the working equations. A feature of the bivariational principle is that the formal bra and ket references can be included as bivariational parameters, which eliminates much of the bias towards the formal reference.  A pilot implementation is described, and extensive benchmark calculations on several standard problems are performed. The results from the bivar-MRCC method are comparable to established state-specific multireference methods. Considering the relative affordability of the bivar-MRCC method, it may become a practical tool for non-experts.
\end{abstract}

\maketitle

\section{Introduction}

In this article, we demonstrate how Arponen's bivariational principle~\cite{Arponen1983} (BIVP)
can be employed to derive a state-specific multireference coupled-cluster (MRCC) method for electronic-structure theory, avoiding many of the problems associated with the currently established state-specific methods, such as sufficiency conditions, non-commuting cluster operators, and so on. The present \emph{proof-of-concept} method is based on single-reference theory, and uses a complete-active space (CAS) approach, but avoids, at least in principle, a bias towards an arbitrary formal reference via an optional bivariational optimization. Thus, the method is not a ``genuine'' multireference method, but should be nearly free of the problems associated with reference bias. We name the method \emph{the bivariational (state-specific) multireference coupled-cluster method} (bivar-MRCC). When reference optimization is included, we name it the \emph{orbital-adaptive} bivariational multireference coupled-cluster method (bivar-OAMRCC). In the same manner as standard single-reference coupled-cluster theory can be viewed as an approximation to Arponen's extended coupled-cluster (ECC) method, we also obtain an  extended version, bivar-(OA)MRECC. The bivariational approach allows a manifestly multiplicatively separable state parameterization, providing automatic extensivity of the energy and computed properties, including excitation energies, whilst being of relative simplicity. Moreover, the bivariational MRCC ansatz should be amenable to relatively straightforward mathematical analysis, e.g., \emph{a priori} error analysis, such as done previously for the ECC method.\cite{Laestadius2018} 
Hence, this approach has the potential of being a powerful quantum chemical tool usable for the non-expert.

Arponen's bivariational approach is top down, starting with potentially \emph{different} but \emph{exact} parameterizations for both a bra and a ket vector $\bra{\tilde{\Psi}}$ and $\ket{\Psi}$. The exact Schrödinger equation is then obtained by requiring the bivariate Rayleigh quotient (expectation value functional) $\braket{\tilde{\Psi}|H|\Psi}/\braket{\tilde{\Psi}|\Psi}$ to be stationary. Approximations are in turn obtained by truncating the state parameters, i.e., a \emph{nonlinear Galerkin approach} in the language of numerical analysis.\cite{Zeidler1990} Mathematical statements of the convergence of the computed results can be made from this top-down approach using basic results from non-linear functional analysis.\cite{Laestadius2018,Zeidler1990} We stress that, while there are ``two wavefunctions'' in bivariational approaches, they form a \emph{unique state} approximation $\rho = \ket{\Psi}\bra{\tilde{\Psi}}/\braket{\tilde{\Psi}|\Psi}$. Since this state is obtained variationally, expectation values are obtained in a straightforward manner using the Hellmann--Feynman theorem.\cite{Feynman1939} Equations for excited states and response theory are also readily formulated.

In his original analysis of CC theory (and the introduction of the ECC method), Arponen used the bivariational approach to write down what was named the \emph{coupled-cluster Lagrangian} by Helgaker and Jørgensen.\cite{Helgaker1989,Helgaker1988} Compared to Arponen's derivation, the conventional CC Lagrangian derivation can be claimed to be bottom up: Starting from the projected similarity-transformed Schrödinger equation, one realizes that its approximate fulfillment via projection is a constrained optimization of the CC energy, and that the corresponding Lagrangian can be conveniently written as an expectation value using an auxiliary bra vector involving the Lagrange multipliers. Thus, in a sense, the bivariational point of view is now \emph{standard} in quantum chemistry, but its power is not fully utilized: the conventional view is very ``ket centric'', while the bivariational top down approach places equal importance to the bra and ket, and the left and right Schrödinger equations, and hence all state parameters. Indeed, for general bivariational methods, the standard notion of a ``projection manifold'' is not meaningful, since the stationary conditions do not decouple bra and ket Schrödinger equations. Finally, let us remark, that all current MRCC theories are similarly focused on the ket side, being based on projections of a similarity transformed Schrödinger equation (or the Bloch equation for state-universal theories). A complete overview of existing state-specific MRCC approaches is beyond the scope of this work. We direct the reader towards the excellent reviews by Köhn \emph{et al.}\cite{Koehn2012}, Lyakh \emph{et al.}\cite{Lyakh2012}, as well as the perspective article by Evangelista.\cite{Evangelista2018}

The bivar-MRCC method resembles the complete-active space coupled-cluster (CASCC) method pioneered by Piecuch, Oliphant, and Adamowicz.\cite{Oliphant1991,Oliphant1992,Piecuch1993} Indeed, the ket ansatz is identical. However, whereas CASCC is based on the projection of the corresponding ket Schrödinger equation, we instead provide an exact bra parameterization. Moreover, the BIVP allows optimization of the formal reference by means of non-orthogonal orbital rotations akin to 
the non-orthogonal orbital-optimized CC (NOCC) method developed by Pedersen
and coworkers.\cite{Pedersen2001} For systems with multireference character, this may lead to significantly more compact wave function representations, in particular
of singlereference type.\cite{Olsen2015,Hiberty2007} Care is taken so that both the bra and the ket vectors are \emph{manifestly} separable, and a balanced treatment of the model space (i.e., the CAS) is obtained for the bra and the ket.

We present first numerical benchmark calculations for the bivar-MRCC and bivar-MRECC methods, performed with a full-configuration interaction (FCI) based pilot implementation. As a multireference method should be be reasonably accurate for both single- and multireference problems, we opted for an example which incorporates both, namely the insertion of Be into H$_2$, a standard example for testing novel multireference coupled-cluster methods since it is computationally feasible even for complicated methods.\cite{purvis83,evangelista11,Lyakh2012} We also perform numerical experiments on the dissociation of the HF and H$_8$ molecules. Whenever possible, we also compare our results with MRCC calculations presented in the literature.

The article is organized as follows: In Section~\ref{sec:bivp} we discuss the BIVP and its Galerkin approximation. We outline how a bivariational method can be analyzed mathematically in order to obtain \emph{a priori} error estimates for the Galerkin approximations. In Section~\ref{sec:mr-ansatz} we introduce the bivar-MRCC method, including the bivariational optimization of the formal reference. We discuss its extensivity and separability properties. In Section~\ref{sec:implementation} we discuss our implementation of the bivar-MRCC method, before we present some numerical results in Section~\ref{sec:numerics}. Finally, in Section~\ref{sec:conclusion} we present our conclusions and future perspectives.

\section{The bivariational principle}
\label{sec:bivp}

A complete mathematical
exposition of the present material is out of scope for the present article, and will be presented elsewhere. The current treatment is compatible
with a finite-dimensional 
Hilbert space $\mathscr{H}$, or alternatively a bounded and below-bounded
Hamiltonian $H$. Virtually all Hamiltonians of interest in molecular
quantum mechanics are unbounded, e.g., they contain a kinetic energy
term. On the other hand, whenever one thinks of a finite-dimensional
full-configuration interaction (FCI) model as ``exact'', the present setting is sufficient.

\subsection{Bivariate Rayleigh quotient}

The BIVP, introduced by Arponen in his seminal
coupled-cluster treatise~\cite{Arponen1983}, and also studied by Löwdin around the same time~\cite{Lowdin1983}, is a generalization of
the Rayleigh--Ritz variational principle to Hamiltonians $H$ that are not
necessarily self-adjoint, even if the most important application is to
these Hamiltonians. The approach introduces, in addition to the usual ket vector $\ket{\Psi}$, the dual vector $\bra{\tilde{\Psi}}$ as a truly 
independent variable, since relaxing the requirement that $H = H^\dag$ makes the left and right eigenvectors independent. The starting point is then the bivariate 
Rayleigh quotient
\begin{equation}
  \mathscr{E}(\tilde{\Psi},\Psi) =
\frac{\braket{\tilde{\Psi}|H|\Psi}}{\braket{\tilde{\Psi}|\Psi}},
\end{equation}
which is stationary if and
only if
\begin{align}
  H\ket{\Psi} = E\ket{\Psi}, \quad \bra{\tilde{\Psi}}H = E
  \bra{\tilde{\Psi}}, \quad \braket{\tilde{\Psi}|\Psi} \neq
  0, \label{eq:se} 
\end{align}
where $E = \mathscr{E}(\tilde{\Psi},\Psi)$. This is the bivariational principle. The basic idea is to introduce potentially \emph{different} approximations to $\bra{\tilde{\Psi}}$ and $\ket{\Psi}$, a flexibility which turns out to be very useful. However, as the bivariate Rayleigh
quotient is not below bounded, unlike the usual variational Rayleigh quotient for a self-adjoint $H$, one cannot throw in any trial bra-ket pair at $\mathscr{E}$
and hope for a meaningful result.

A potentially confusing aspect of the BIVP is the fact that we now have ``two wavefunctions''. However, the \emph{state} formed is unique, i.e., it is a non-Hermitian rank-one density operator $\rho = \ket{\Psi}\bra{\tilde{\Psi}}/\braket{\tilde{\Psi}|\Psi}$. Since $\rho$ is determined variationally, the Hellmann--Feynman approach\cite{Feynman1939} can be used to define expectation values of arbitrary observables $A$, i.e., 
\begin{equation}
    \braket{A} \equiv \operatorname{Tr} \rho A = \frac{\braket{\tilde{\Psi}|A|\Psi}}{\braket{\tilde{\Psi}|\Psi}}. \label{eq:bivp-expt-val}
\end{equation}
By introducing the \emph{time-dependent} BIVP~\cite{Arponen1983,Chernoff1974}, we can take the leap to the time domain. The bra and ket time-dependent Schrödinger equations are obtained as stationary points of the action-like integral
\begin{equation}
    \mathscr{S} =     \int_0^T \bra{\tilde{\Psi}(t)}(i \partial_t - H) \ket{\Psi(t)} \, dt.
\end{equation}
This opens up the route to not only response theory\cite{Koch1990} and the approximation of excited states\cite{Arponen1983,Arponen1987b}, but also real-time propagation of quantum systems far from the ground-state\cite{Kvaal2012,Sato2018,Pedersen2019,Kristiansen2020}.

\subsection{Parameterization maps and discretization}

Suppose that we are given a parameterization map
$\chi : \tilde{V}\oplus V \to \tilde{\mathscr{H}}\oplus\mathscr{H}$,
where $V$ is some Hilbert space, and where $\tilde{\mathscr{H}}$
(resp.~$\tilde V$) is the dual space of $\mathscr{H}$ (resp.~$V$), i.e., space of bra vectors. The map $\chi$ is
assumed to be smooth and with a smooth inverse near a ground-state
pair $(\bra{\tilde\Psi_*},\ket{\Psi_*})$, i.e., the parameterization is \emph{exact} near the ground state. The map induces an energy
expectation value functional
$\mathscr{E}_\chi : \tilde{V} \oplus V \to \mathbb{C}$, with $\mathscr{E}_\chi(\tilde{v},v) = \mathscr{E}(\tilde{\Psi}(\tilde{v},v),\Psi(\tilde{v},v))$, which is smooth, and
whose critical points are in one-to-one correspondence
with the solutions of Eq.~\eqref{eq:se} that can be reached with
$\chi$. In particular, the ground state is parameterized by a point $(\tilde{v}_*,v_*) \in \tilde{V}\oplus V$. It follows that the Schr\"odinger equation and its dual can be
written: Find $(\tilde{v}_*,v_*)\in\tilde{V}\oplus V$ such that
\begin{equation}
  \frac{\partial}{\partial \tilde{v}} \mathscr{E}_\chi(\tilde{v}_*,v_*) = 0, \quad 
  \frac{\partial}{\partial v} \mathscr{E}_\chi(\tilde{v}_*,v_*) = 0. \label{eq:se2}
\end{equation}
A bivariational approximation is now obtained by a Galerkin approach defined by projection in the
space $V$, i.e., we restrict Eq.~\eqref{eq:se2} to the space
$\tilde{V}_d\oplus V_d$, where the subscript $d$ is a discretization parameter: Find $(\tilde{v}_{d*},v_{d*}) \in \tilde{V}_d\oplus V_d$ such that 
\begin{equation}
    P_{V_d} \frac{\partial}{\partial \tilde{v}_d} \mathscr{E}_\chi(\tilde{v}_{d*},v_{d*}) = 0, \quad  P_{{\tilde V}_d} \frac{\partial}{\partial v_d} \mathscr{E}_\chi(\tilde{v}_{d*},v_{d*}) = 0.\label{eq:se2-p}
\end{equation}
Here, $P_{V_d}$ is the projector onto $V_d$.
In any Galerkin approach, it is assumed that any $v\in V$ can be approximated
sufficiently well by the projections $v_d = P_{V_d} v \in V_d$, i.e.,
$\|v_d - v\|\to 0$ as $d\to \infty$, (and similarly for the dual element). The example to keep in mind is that of 
$V$ being the space of cluster amplitudes (or operators), and $V_d$ being a  
a single (S), double (D), etc., truncation. The limit $d\to \infty$
corresponds to the untruncated limit (and also the basis set limit in the infinite dimensional case). Typically,  $V$ consists of excitation amplitude vectors, and $\tilde{V}$ consists of de-excitation amplitude vectors.

\subsection{Local strong monotonicity analysis}

Several questions arise: First, does the discrete bivariational
Schr\"odinger equation~\eqref{eq:se2-p} have a solution? Is this solution unique? Does
the solution $(\tilde{v}_{d*},v_{d*})$ and the corresponding energy
$E_d=\mathscr{E}_d(\tilde{v}_{d*},v_{d*})$ converge to the exact solution
$(\tilde{v}_*,v_*)$ and energy $E_* = \mathscr{E}_\chi(\tilde{v}_*,v_*)$, respectively?

This problem was analyzed and under mild sufficient conditions, the
questions answered in the affirmative by Laestadius and Kvaal for
Arponen's ECC method~\cite{Laestadius2018,Kvaal2020},
implying the same results for the standard coupled-cluster
method. For an analysis of standard CC without the BIVP, see the works of
Rohwedder and Schneider~\cite{Rohwedder2013,Rohwedder2013b}. In Section~\ref{sec:mr-ansatz}, we will introduce our multireference ansatz, i.e., our map $\chi$. However, we will relegate the mathematical analysis of the method to future work. See however Ref.~\onlinecite{Faulstich2019}, where Faulstich \emph{et al.} studied Kinoshita's tailored coupled-cluster method\cite{Kinoshita2005}. Since this method also uses a CAS model space, it is likely that the assumptions needed for the analysis of bivar-MR(E)CC will be similar. Even if the mathematical analysis is postponed, we find it instructive to explain its basic ingredients, which should serve as compelling evidence for the usefulness of the bivariational approach in general.

A
key analysis tool for studying CC theory~\cite{Laestadius2018,Laestadius2019,Rohwedder2013,Rohwedder2013b,Faulstich2019,Kvaal2020} was that of local strong monotonicity\cite{Zeidler1990} of the \emph{flipped
gradient} $F : \tilde{V}\oplus V \to \tilde{V}'\oplus V'$ given by
\begin{equation}
  F(\tilde{v},v) = \left( \frac{\partial}{\partial
      v}\mathscr{E}_\chi(\tilde{v},v), \frac{\partial}{\partial
      \tilde{v}}\mathscr{E}_\chi(\tilde{v},v) \right). \label{eq:flipped}
\end{equation}
Local strong monotonicity can be rephrased as the Jacobian of $F$
being coercive at the ground state. Flipping the gradient is motivated by the following: The simple gradient of the original bivariate Rayleigh quotient $\mathscr{E}$ is \emph{not} monotone, since every eigenvalue is a saddle point. However, it can be demonstrated that flipping the gradient turns the ground-state saddle point into something like a local minimum under reasonable conditions on the Hamiltonian. Thus, it makes sense to consider Eq.~\eqref{eq:flipped} and find conditions on $\chi$ such that local strong monotonicity is inherited. 

For such an analysis, it is easiest to work with a parameterization in which the normalizations of the bra and ket are fixed. In the setting of ECC, the bra-ket pair is normalized
according to $\braket{\tilde\Psi|\Psi}=\braket{\Phi_0|\Psi}=1$, where
$\ket{\Phi_0}$ is the reference determinant in single-reference
theory. 

When $F$ is locally strongly monotone near the ground-state,
Zarantonello's Theorem~\cite{Zeidler1990,Laestadius2018,Kvaal2020} on local form implies
$v_{d*}\to v_*$ and $\tilde{v}_{d*} \to \tilde{v}_*$ as $d\to \infty$. The critical point formulation of the
Schr\"odinger equation immediately implies a quadratic error estimate,
\begin{equation}
  |E_* - E_d| \leq C ( \|v_*-v_{d*}\|^2 + \|\tilde{v}_*-\tilde{v}_{d*}\|^2 ),
\end{equation}
for some constant $C$, which holds for sufficiently large $d$, i.e., for sufficiently large Galerkin subspaces $V_d$. We remark, that this is a \emph{local} result. There may be solutions of the discrete equations that are not related to exact solutions, and the local convergence is only achieved sufficiently far into the Galerkin sequence of spaces, i.e., for large enough $d$.

\section{State-specific multireference formulation}
\label{sec:mr-ansatz}

\subsection{Bra and ket model spaces}

\begin{figure}[htb]
    \centering
\includegraphics[]{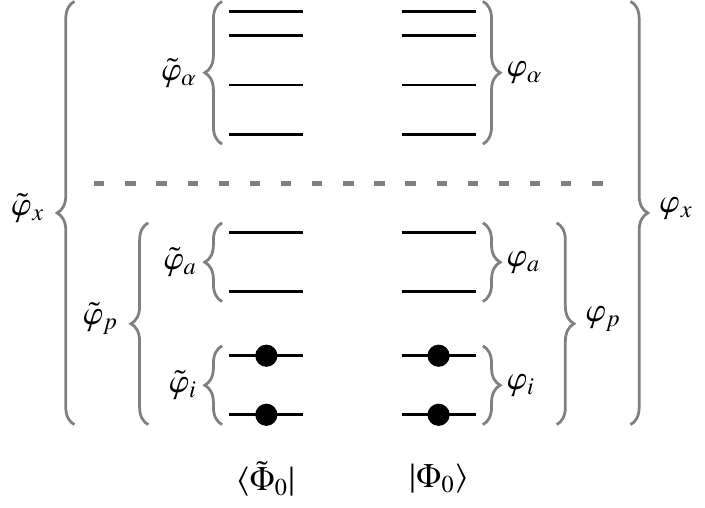}
    \caption{Illustration of the composition of the bra and ket single-particle spaces in terms of biorthogonal spin-orbitals, including the index conventions used in this work. The dashed line indicates the end of the CAS. The bra and ket formal references are shown.}
    \label{fig:orbitalspace}
\end{figure}

Like all multireference methods, we need a specification of a \emph{model space} $\mathscr{H}_0$ and an \emph{external space} $\mathscr{H}_\text{ext}$, together forming the \emph{computational} $N$-electron space $\mathscr{H} = \mathscr{H}_0 \oplus \mathscr{H}_\text{ext}$. The computational space is often a proper subspace of the full $N$-electron space $\mathscr{H}_{N}$. For simplicity, we take as model space a standard CAS generated by a finite set of single-particle functions (spin-orbitals) $\{\varphi_p\} = \{\varphi_i\} \cup \{\varphi_a\}$, divided into occupied $\{\varphi_i\}$ and unoccupied $\{\varphi_a\}$ subsets (Fig.~\ref{fig:orbitalspace}). For a general CAS, the occupied orbitals are again divided into active and inactive subsets, while $\varphi_a$ is always denoted active. The formal reference $\ket{\Phi_0}$ is composed of the occupied $\{\varphi_i\}$. Finally, the external space is obtained by choosing another set  $\{\varphi_\alpha\}$ of virtual functions, linearly independent to $\{\varphi_p\}$, and taking as determinantal basis for $\mathscr{H}_{\text{ext}}$ the determinantal basis for $\mathscr{H}_0$, with at least one substitution of an active single-particle function by an external single-particle function $\varphi_\alpha$.

The whole computational space is now generated by the set of spin-orbitals $\{\varphi_x\} = \{\varphi_i\}\cup\{\varphi_a\}\cup\{\varphi_\alpha\}$. We do \emph{not} assume that this set is orthonormal. Instead, we will assume biorthogonality with the dual spin-orbitals that define the dual computational Hilbert space $\tilde{\mathscr{H}} = \tilde{\mathscr{H}}_0\oplus \tilde{\mathscr{H}}_\text{ext}$. To this end, we introduce dual active spin-orbitals $\{\tilde{\varphi}_x\} = \{\tilde{\varphi}_i\}\cup\{\tilde{\varphi}_a\}\cup\{\tilde{\varphi}_\alpha\}$ that generate $\tilde{\mathscr{H}}_0$ and $\tilde{\mathscr{H}}_\text{ext}$, in the same manner as above. Biorthogonality means  $\braket{\tilde{\varphi}_x|\varphi_y} = \delta_{xy}$. The dual reference $\bra{\tilde{\Phi}_0}$ is defined by occupying all the $\tilde{\varphi}_i$, from which we obtain $\braket{\tilde{\Phi}_0|\Phi_0} = 1$. Indeed, we have in general $\braket{\tilde{\Phi}_\mu|\Phi_\nu} = \delta_{\mu\nu}$, where $\mu$ is a generic numbering scheme of the Slater determinant basis of the computational Hilbert space.

A general model space ket can be written
\begin{equation}
  \ket{\Psi_0} = \sum_{\mu\in\text{CAS}} \ket{\Phi_\mu} c_\mu, \label{eq:msket}
\end{equation}
where $\mu\in\text{CAS}$ indicates that the sum is over the CAS determinant basis only. Similarly, a general model space bra can be written
\begin{equation}
  \bra{\tilde{\Psi}_0} = \sum_{\nu\in\text{CAS}} d_\nu \bra{\tilde{\Phi}_\nu}.\label{eq:msbra}
\end{equation}
We note that $\braket{\tilde{\Psi}_0|\Psi_0} =  \sum_{\mu\in\text{CAS}} d_\mu c_\mu = d^T c$, introducing matrix notation for the amplitudes.

Instead of a CAS, we may also consider incomplete active spaces or even more general model spaces and external spaces. For the following, the only really important feature is that \emph{the model space is spanned by determinants that are excitations from the formal reference}, and that \emph{the external space is generated by excitations out of the model space}.

As outlined in the Introduction, we intend to allow the formal references to be optimized bivariationally, i.e., the \emph{active} occupied $\tilde{\varphi}_i$ and $\varphi_i$ are variables in the parameterization map $\chi$ to be described. For the moment, however, we consider these active occupied spin-orbitals to be fixed.

\subsection{Bra and ket parameterizations in Hilbert space}

In the internally contracted MRCC scheme\cite{Koehn2012,Lyakh2012,Evangelista2018} (ic-MRCC), a general wavefunction
$\ket{\Psi}\in \mathscr{H}$ is written
\begin{equation}
  \ket{\Psi} =  e^{T_\text{ic-MRCC}} \ket{\Psi_0},\label{eq:ic-mrcc}
\end{equation}
where $T_\text{ic-MRCC} = \sum_\mu T^{(\mu)}$ is a general cluster operator, whose components $T^{(\mu)}$ are single-reference cluster operators relative to $\ket{\Phi_\mu}$ as reference.
We note that $T^{(\mu)}$
is in general not unique, and that $[T^{(\mu)},T^{(\nu)}]\neq 0$. These are among the basic problems which we would like to avoid.

Suppose now that $\ket{\Psi} \in \mathscr{H}$ has a nonzero component along the formal reference
$\ket{\Phi_0}$. We can then use single-reference CC theory to \emph{uniquely} write
$\ket{\Psi} = \exp(T_\text{SR})\ket{\Phi_0}$, where
$T_\text{SR}=T_0 + T$ is the full cluster operator in single-reference
theory manner. The term $T_0$ contains precisely those excitations
that stay within the model space, leaving $T$ as the external
excitations having at least one external label $\alpha$. Exploiting
this, we obtain 
\begin{equation}
    \ket{\Psi} = e^T \ket{\Psi_0}, \label{eq:cascc-ket}
\end{equation}
where $\exp(T_0)$ has been
converted to a FCI expansion. The operator $T$ is
unique, since any external space determinant is uniquely obtained as an
excitation from $\ket{\Phi_0}$. 

Similar to the above considerations, a general bra $\bra{\Omega}$ with nonzero component along $\bra{\tilde{\Phi}_0}$ can be written $\bra{\Omega} = \bra{\tilde{\Psi}_0} e^S$, where $S$ is an external de-excitation operator (excitation operator
for bras).  In the spirit of Arponen's ECC method, we can postmultiply with the invertible operator $e^{-T}$ to obtain the bra parameterization
\begin{equation}
  \bra{\tilde{\Psi}} = \bra{\tilde{\Psi}_0} e^Se^{-T} , \quad
  \bra{\tilde{\Psi}_0} = \sum_{\mu\in\text{CAS}} d_\mu \bra{\tilde{\Phi}_\mu}. \label{eq:tilde-psi}
\end{equation}
This is valid so long as $\bra{\tilde{\Psi}}e^T\ket{\Phi_0} \neq 0$, a very mild restriction.
We note that $\braket{\tilde{\Psi}|\Psi} =
\braket{\tilde{\Psi}_0|\Psi_0} = d^T c$. 

This now completes the specification of an \emph{exact} parameterization map $(\bra{\tilde{\Psi}},\ket{\Psi}) = \chi(s,t,d,c)$, where $s$ and $t$ are the amplitudes of $S$ and $T$, respectively. Here, the dependence on  $\{\varphi_x\}$ and $\{\tilde{\varphi}_x\}$ is implicitly included via the determinantal basis' dependence on these. Plugging into the bivariate
Rayleigh quotient, we obtain the energy
functional of bivar-MRECC,
\begin{equation}
\begin{split}
  &\mathscr{E}_\text{bivar-MRECC}(s,t,d,c) \\& \qquad\quad= \braket{\tilde{\Phi}_0|DC|\Phi_0}^{-1} \braket{\tilde{\Phi}_0|D e^S e^{-T} H e^T C|\Phi_0},
 \end{split}
\end{equation}
where we introduced the CAS cluster operators $C$ and (de-excitation) operator $D$. 
We remark, that in Arponen's ECC, a further coordinate change $(t,s) = (t(s,s'),s)$ is made, where $s'$ are the amplitudes of a cluster operator $S'$ defined by $\braket{\Phi_\mu|S'|\Phi_0} = \braket{\Phi_\mu| e^{S} T|\Phi_0}$. This is done in order to ensure a certain linkedness structure of the diagram series, and also has the implication that the \emph{time-dependent} Schrödinger equation takes the form of a canonical Hamiltonian system.\cite{Arponen1983,Arponen1987,Kvaal2020} We will not further explore the ECC flavor of the multireference ansatz here, but instead reparameterize $\Lambda = e^S-1$ to obtain the energy functional of the bivar-MRCC method,
\begin{equation}\label{eq:bivar-MRCC-energy}
\begin{split}
  &\mathscr{E}_\text{bivar-MRCC}(\lambda,t,d,c)  \\
  & \qquad  = \braket{\tilde{\Phi}_0|DC|\Phi_0}^{-1} \braket{\tilde{\Phi}_0|D (1 + \Lambda) e^{-T} H e^T C|\Phi_0}
  \\
  &\qquad =  (d^T c)^{-1} d^T  K(t,\lambda) c, 
  \end{split}
\end{equation}
where $K(t,\lambda) = [K(t,\lambda)_{\mu\nu}]=[\braket{\tilde{\Phi}_\mu|(1+\Lambda)e^{-T}He^T|\Phi_\nu}]$ can be considered an effective CAS
Hamiltonian.

Finally, we consider the special case when the CAS has a nonempty set of \emph{inactive} occupied spin-orbitals. Let $\mathscr{H}_{0,N}$ be the space spanned by \emph{all} determinants built from $\{\varphi_p\}$, such that the \emph{full} $N$-electron space is divided as $\mathscr{H}_N = \mathscr{H}_{0,N} \oplus \mathscr{H}_{0,N}^\bot$. Clearly, $\mathscr{H}_0\subset \mathscr{H}_{0,N}$ and $\mathscr{H}_{\text{ext}} \subset \mathscr{H}_{0,N}^\bot$ are proper whenever there are inactive occupied orbitals. In this case, a general single-reference cluster operator $T_\text{SR} = T_0 + T$ has components that generate Slater determinants that violate the occupancy restrictions in the CAS and the external space,\cite{Silverstone1966,fink93,doi:10.1063/1.455556,doi:10.1063/1.5040587} i.e., these determinants are in $\mathscr{H}_{N}$ but not in $\mathscr{H}$. Thus, $T$ is not freely variable, but must be constrained. We treat this as a technical problem to be addressed in the numerical implementation, see Sec.~\ref{sec:implementation-amplitude-projection}. In the remainder of Sec.~\ref{sec:mr-ansatz}, we therefore assume that there are no inactive occupied spin-orbitals, so that $T$ and $\Lambda$ are not constrained.

\subsection{Truncation schemes}
\label{sec:truncations}

We briefly consider Galerkin schemes for the bivar-MR(E)CC method, i.e., cluster operator truncations. The model space expansion coefficients $c$ and $d$ are in this work never truncated. The untruncated cluster operators read
\begin{equation}
    T = \sum_{\mu\in\text{ext}} t_\mu X_\mu, \quad \Lambda = \sum_{\mu\in\text{ext}} \lambda_\mu Y_\mu,
\end{equation}
where $\mu\in\text{ext}$ denotes a general external excitation index. We employ the usual truncation scheme in terms of external singles (S), doubles (D), etc., relative to the formal reference. The result is the SD\ldots$K(n,m)$ truncation scheme, built from a CAS with $n$ electrons in $m$ (spatial) orbitals, and external single and double excitations, etc., up to $K$-fold excitations.

In order to obtain a more balanced description for all model space states, 
in particular when degeneracies are present, it might be necessary that the cluster operators include excitations out of all model space determinants.\cite{doi:10.1063/1.481649} The simplest choice is the first-order interaction space (FOIS)\cite{Roosbook,Lyakh2012,doi:10.1063/1.481649} defined by all single and double excitations relative to the \emph{model space} into the external space. Inclusion of the FOIS in the cluster operators ensures that computed energies will be correct through second order in perturbation theory. This is so, because the excitations in the FOIS are precisely those that are coupled to the model space $\mathscr{H}_0$ by two-body Hamiltonians. Thus, the FOIS consists of selected external doubles, triples, quadruples, and so on. The resulting truncation will be denoted SD\ldots $K(n,m)$FOIS. We remark that the same approach is employed in the CASCC method and also reflects the excitation manifold used in internally contracted multireference methods.

\subsection{Working equations}
\label{sec:working-equations}
We proceed to discuss the stationary conditions of the bivar-MRCC energy. The equations for the extended version are similarly obtained, and omitted here.

Differentiation of Eq.~\eqref{eq:bivar-MRCC-energy} with respect to the CAS amplitudes $d_\mu$ and $c_\mu$ yields, respectively, right and left eigenvalue equations for the effective Hamiltonian matrix $K = K(t,\lambda)$,
\begin{subequations}
\label{eq:bivar-MRCC-working}
\begin{equation}
    Kc = Ec, \quad \text{and}\quad K^T d = E d, \label{eq:bivar-MRCC-working-evp}
\end{equation}
as well as the condition $d^Tc \neq 0$. Here, $E=E(t,\lambda)=\mathscr{E}_\text{bivar-MRCC}(\lambda,t,d,c)$ can, in the regime of weak dynamical correlation, be taken to be the smallest eigenvalue. However, in practice the ground-state solution may correspond to a higher eigenvalue; see Sections~\ref{sec:implementation} and \ref{sec:numerics}. Without loss, we can assume that $d^T c = 1$ at the solution.

Differentiation with respect to $\lambda_\mu$ gives a ($\lambda$-independent) equation for $t$, 
\begin{equation}
    \Omega_\mu(t,d,c) := \braket{\tilde{\Phi}_\mu| D e^{-T} H e^T C |\Phi_0} = 0. \label{eq:bivar-MRCC-working-t}
\end{equation}
Finally, differentiation with respect to $t_\mu$ gives a linear equation for $\lambda$,
\begin{equation}\label{eq:bivar-MRCC-working-lambda}
\begin{split}
    \tilde{\Omega}_\mu(\lambda,t,d,c) :=& \braket{\tilde{\Phi}_0|D [e^{-T}H e^T, X_\mu] C|\Phi_0} \\&+ \sum_{\nu\in\text{ext}} \braket{\tilde{\Phi}_\nu|D [e^{-T}H e^T, X_\mu] C|\Phi_0}\lambda_\nu = 0. 
    \end{split}
\end{equation}
\end{subequations}
The $t$-equations~\eqref{eq:bivar-MRCC-working-t} and the $\lambda$-equations~\eqref{eq:bivar-MRCC-working-lambda} are similar in structure as the corresponding equations in standard singlereference CC theory.

\subsection{Bivariational optimization of reference}
\label{sec:ref-opt}

We now describe the orbital-adaptive element of the bivar-MRCC method, i.e., the bivar-OAMRCC formulation. In order to alleviate the arbitrariness of the formal bra and ket reference determinants,
we introduce optional orbital rotations in the model space as bivariational
parameters, i.e., we let the active occupied orbitals $\varphi_i$ and
$\tilde{\varphi}_i$ be variational parameters in $\mathscr{E}_\text{bivar-MRCC}$.
The spin-orbitals are free to vary within the given single-particle
model space. Since the singlereference CC ansatz is invariant under
separate rotations of occupied and virtual orbitals, it is sufficient
to consider orbital variations of the form
\begin{equation}
  \varphi_p \longrightarrow \sum_q \varphi_q (e^{\kappa})_{qp},\quad\text{and}\quad
  \tilde{\varphi}_q \longrightarrow \sum_p (e^{-\kappa})_{qp}
                      \tilde{\varphi}_p ,\label{eq:bivar-MRCC-working-orb-transform}
\end{equation}
where $\kappa=[\kappa_{pq}]$ is a non-singular matrix with $\kappa_{ij}=
\kappa_{ab} = 0$. The transformation preserves biorthogonality of the
single-particle functions.

The determinants transform as
\begin{equation}
  \ket{\Phi_\mu} \longrightarrow e^{\hat{\kappa}} \ket{\Phi_\mu},\quad\text{and}\quad
  \bra{\tilde{\Phi}_\mu} \longrightarrow \bra{\tilde{\Phi}_\mu} e^{-\hat{\kappa}},
\end{equation}
with
\begin{equation}
  \hat{\kappa}  = \sum_{ia} \kappa_{ai} c^\dag_a \tilde{c}_i -
  \kappa_{ia} c^\dag_i \tilde{c}_a \equiv \hat{\kappa}_+ -
  \hat{\kappa}_- .
\end{equation}
Here, $\tilde{c}_q$ is the destruction operator associated with the
dual spin-orbital $\tilde{\varphi}_q$, and $c^\dag_p$ is the creation operator associated with $\varphi_p$.
The non-zero matrix elements of $\kappa$ are all independent, and we
can express the energy functional in terms of $\kappa$, given an arbitrary fixed
``guess'' of orbitals, viz.,
\begin{equation}
\begin{split}
&\mathscr{E}_\text{bivar-OAMRCC}(\lambda,t,d,c,\kappa_-,\kappa_+) \\ &\quad  =  \braket{\tilde{\Phi}_0|DC|\Phi_0}^{-1}
\braket{\tilde{\Phi}_0|D(1 + \Lambda)e^{-T}e^{-\hat\kappa} H e^{\hat\kappa} e^{T}
  C|\Phi_0}.
\end{split}
\end{equation}
Assuming that the current basis is actually the solution,
i.e., $\kappa=0$ is the critical point, we obtain stationary
conditions from the first-order term of the Baker–Campbell–Hausdorff (BCH) series of
$e^{-\hat{\kappa}}He^{\hat{\kappa}}$,
\begin{subequations} \label{eq:bivar-MRCC-working-orb}
\begin{align}
  0 &=  \braket{\tilde{\Phi}_0|D[ (1
    + \Lambda)e^{-T}H e^{T}, c^\dag_a \tilde{c}_i] C| \Phi_0} ,
  \\
  0 &=   \braket{\tilde{\Phi}_0|D[ (1
    + \Lambda)e^{-T}H e^{T}, c^\dag_i \tilde{c}_a] C | \Phi_0},
\end{align}
\end{subequations}
for every pair of $(a,i)$. 

It should be noted that, in the full, untruncated bivar-MRCC model, \emph{all} the matrix elements of $\kappa$ are redundant, since we are at the FCI limit. Thus, reference optimization only makes sense in a truncated bivar-MRCC model. Moreover, in the case where the truncation leads to an accurate dynamical correlation representation, it can be expected that the orbital dependence is weak.

\subsection{Extensivity}
\label{sec:discussion}

In the extended bivar-MRECC parameterization, \emph{both} the bra and the ket are \emph{manifestly} multiplicatively separable for noninteracting subsystems, so long as the model space is  complete and a FCI expansion is kept. (For incomplete active spaces and general model spaces the analysis is more involved.\cite{Nooijen2005}) 
It follows that the energy functional is additively separable (extensive), and that expectation values and properties are also extensive. Excitation energies computed using linearization of the equations of motion (equivalently, response theory) yields intensive excitation energies, also for independent excitations on each subsystem. This should be contrasted to standard equation-of motion coupled-cluster theory (EOM-CC), where such excitations are not additive.\cite{PinkBible} This can be traced to the bra not being multiplicatively separable.

For the linear bivar-MRCC version, the bra is not fully multiplicatively separable due to linearity in $\Lambda$. However, the state \emph{is} separable in the CAS part. Thus, we expect excitation energies to be very well represented, also for individual excitations on noninteracting subsystems, as long as the model space resolves the system's quasidegeneracy.

It is instructive to compare the bivar-MR(E)CC method to the CASCC method of Adamowicz and coworkers, which can be defined in terms of the Lagrangian-like functional
\begin{equation}
    \begin{split}
    &\mathscr{E}_{\text{CASCC}}(\lambda,t,d,c) \\ & \quad = \braket{{\Phi}_0|DC|\Phi_0}^{-1} \braket{{\Phi}_0|(\Lambda + D)e^{-T}He^T C|\Phi_0}.
    \end{split}
\end{equation}
This Lagrangian, where orthonormal spin-orbitals are assumed, is derived from the projection of the similarity transformed Schrödinger equation $e^{-T}He^T C \ket{\Phi_0} = E C \ket{\Phi_0}$.  While the ket is identical to bivar-MRCC, and hence separable, separability of the bra is lost. Indeed, whereas the bivar-MRCC method has an (approximate) multiplicative structure in the bra, the CASCC bra has no such structure, resulting in two quite different methods. While the CASCC method is cheaper and more straightforward to implement, we conjecture that the multiplicative structure of the bivar-MR(E)CC methods will have strong impact on the  computed properties, response theory, and in particular excitation energies.

\section{Implementation}
\label{sec:implementation}

An implementation of the working equations~\eqref{eq:bivar-MRCC-working} for the bivar-MRCC and bivar-MRECC methods requires solving a non-symmetric CI, a CC-type and, in the case where orbital-adaptivity is included via Eq.~\eqref{eq:bivar-MRCC-working-orb}, a mean field problem. All these are coupled. The simplest approach is an iterative approach, where these subproblems are solved in turn.

In our pilot implementation, determinant based FCI technology is used, i.e., wave functions are expanded in a FCI basis, and matrix elements over general excitation operators are decomposed into contributions of type $\langle \tilde{\Phi}_\mu | \cramped{c_{x}^\dag} \tilde{c}_y | \Phi_\nu\rangle$ by inserting the FCI identity.\cite{KNOWLES198975}  All operations on FCI vectors are SMP-parallelized, and expectation values are computed by evaluating inner products.  For simplicity, spin-symmetry is exploited only in the computation of matrix-vector products involving the Hamiltonian (also known as $\sigma$-vectors).  Up to four vectors of full length are kept in memory, thus limiting the scope of applicability to performing benchmark studies on small model systems. A more efficient implementation is currently in development, and uses the fact that singlereference technology can be used to compute the contributions that feature the same model space determinant (generated by $D$ and $C$ operators) on the left and right hand side of Eq.~\eqref{eq:bivar-MRCC-working} and Eq.~\eqref{eq:bivar-MRCC-working-orb}, respectively. This approach has been discussed for Mukherjee's state-specific Mk-MRCCSD method~\cite{Prochnow2009} and leads to an effective scaling of  $\mathcal{O}(n_\text{MS} n_{\text{occ}}^2 n_{\text{virt}}^4)$ for these terms, with $n_\text{MS}$ being the number of model space determinants, and $\mathcal{O}(n_{\text{occ}}^2 n_{\text{virt}}^4)$ the scaling of the conventional CCSD method.  The remaining contributions are very sparse and automated code generation together with tensor contraction technology can be used to compute these efficiently.~\cite{Hirata2003,Solomonik2014,Schutski2017} 

The external cluster operators used in the present implementation are defined by the SD\ldots $K(n,m)$ and SD\ldots $K(n,m)$FOIS schemes, see~Section~\ref{sec:truncations}. A collective $K$-fold excitation index $\mu\in\text{ext}$ is represented by the $2K$-tuple
\begin{equation}
    \begin{split}\label{eqn:impl-exc-tuple}\vphantom{\int}
     \mu \to (\iota_1^{{(\mu)}}\!\!\!,\dots,\iota_M^{(\mu)}\!,&i_{M+1}^{(\mu)},\dots,i_{K}^{(\mu)}\!, \\ & a_1^{(\mu)}\!\!\!,\dots a_{K-M'}^{(\mu)},\alpha_{K-M'+1}^{(\mu)},\dots,\alpha_K^{(\mu)}),
    \end{split}
\end{equation}
with $\cramped{\iota_1^{(\mu)}}>\dots > \cramped{\iota_M^{(\mu)}}$, $\cramped{i_{M+1}^{(\mu)}}>\dots >\cramped{i_{K}^{(\mu)}}$, $\cramped{a_1^{(\mu)}}> \dots > \cramped{a_{K-M'}^{(\mu)}}$, and $\cramped{\alpha_{K-M'+1}^{(\mu)}}> \dots >\cramped{\alpha_K^{(\mu)}}$. In Eq.~\eqref{eqn:impl-exc-tuple}, inactive occupied orbitals are counted using $\iota$ and the internal amplitudes (defined by $M=M'=0$) are excluded.
For a general CAS with inactive orbitals (i.e., $M>0$),
linear dependencies  (e.g., induced by double excitations containing active-active ``spectator excitations'') are removed efficiently by orthogonalization (\emph{vide infra}). In each step of the mean-field optimization Eq.~\eqref{eq:bivar-MRCC-working-orb}, the integrals with active indices are transformed according to Eq.~\eqref{eq:bivar-MRCC-working-orb-transform},
ensuring that $X_\mu$ is (Hermitian) adjoint to $Y_\mu$ at any step in the computation.

In our current implementation, the maximum excitation rank is not limited, thus in principle allowing for arbitrary order cluster operators. Excitations from the model space to the FOIS are implemented by redefinition in terms of  excitations with respect to the reference determinant. Thus, a bivar-MR(E)CCSD($n$,$m$)FOIS cluster operator contains maximally $(n+2)$-fold excitations with respect to the reference determinant.\cite{Lyakh2012}

\begin{figure}[htb]
\includegraphics[]{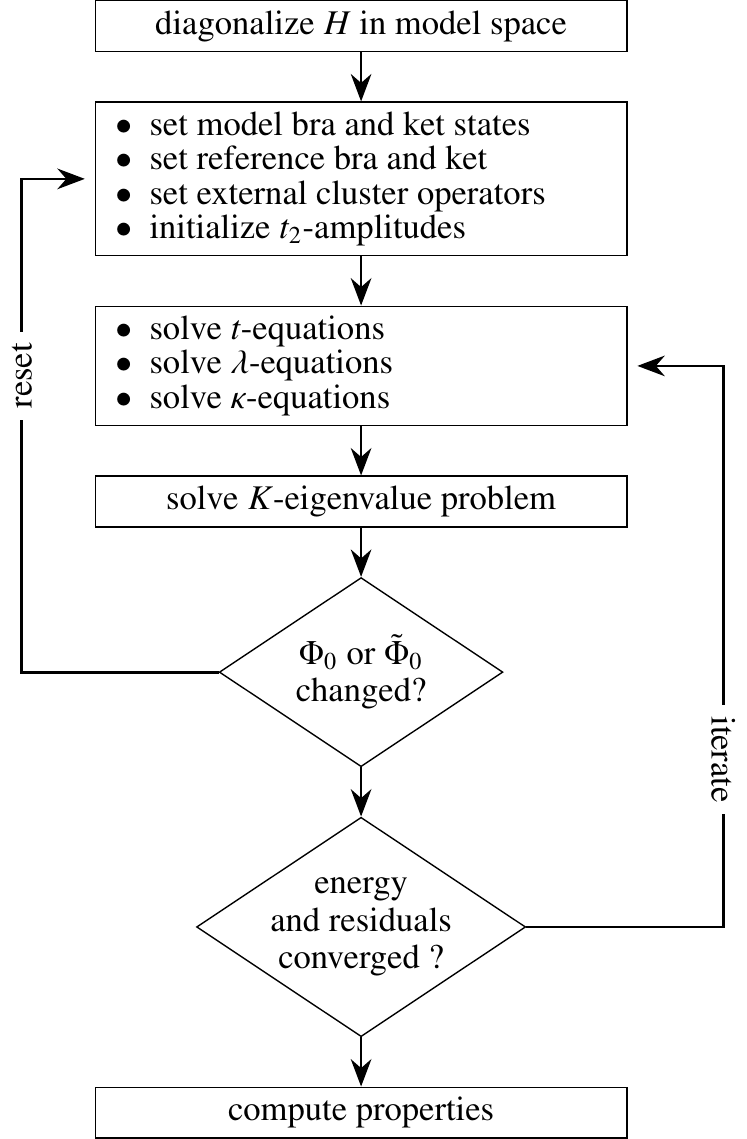}
\caption{Iteration scheme for a bivar-MR(E)CC computation. After the initial diagonalization of $H$ in the CAS model space and the consequent initialization of bivar-MR(E)CC variables, the main loop solves the various working equations in turn. If the reference changes after the $K$-matrix diagonalization, the bivar-MR(E)CC variables need to be reset. The iteration truncates after the energy changes less than a given tolerance and the residual norms are sufficiently small. See text for further details.\label{fig:compscheme}}
\end{figure}

The computation of the bivar-MR(E)CC wavefunctions and energies is performed iteratively (see Fig.~\ref{fig:compscheme}): First, based on a (fixed) model space definition, the Hamiltonian is diagonalized in this subspace to give the model-space wave functions~\eqref{eq:msket} and \eqref{eq:msbra}. Based on these CASCI vectors, the (initial) reference state and reference determinant are defined. Using this definition, the doubles part of the $t$-amplitude vector is populated with second-order Møller--Plesset (MP2) values. Then, the $t$- and $\lambda$-equations are solved iteratively. In the case of bivar-MRECC, $t$ and $\lambda$ are optimized simultaneously. If orbital adaptivity is considered, the orbitals are optimized either before or after solving the CC problem by solving Eq.~\eqref{eq:bivar-MRCC-working-orb}, and the integrals are transformed into a basis where all ${\kappa}_{pq}=0$.\cite{doi:10.1063/1.5006160} Finally, the matrix $K(t,\lambda)$ (Eq.~\eqref{eq:bivar-MRCC-energy}) is constructed within the model space and diagonalized to give the updated model space vectors $c$ and $d$. The amplitude vectors $t$ and $\lambda$ and the CI coefficients are re-optimized until convergence, which is typically achieved in 3 to 10 (outer) iterations.

\label{sec:reference-wave-function-during-iteration}
During iterations, the character of the reference wave functions Eq.~\eqref{eq:msket} and Eq.~\eqref{eq:msbra} is preserved by choosing those eigenvectors of Eq.~\eqref{eq:bivar-MRCC-working-evp} that have the largest overlap with the $c$ and $d$ vectors from the previous iteration thereby avoiding problems created by root flipping. Thus, the initial choice immediately after the CASCI step defines the nature of the state that is optimized. Similarly to the CASCC(sw) method~\cite{adamowicz15}, the reference determinant is allowed to change dynamically during iterations. Whenever this happens, the definition of the cluster operators is reset and the amplitudes are reinitialized to MP2 values.

\label{sec:implementation-amplitude-projection}

Convergence of the $t$- and $\lambda$-equations is accelerated by using a quasi-Newton-Raphson update together with the direct inversion in the iterative subspace technique.\cite{PULAY1980393} If the set of inactive orbitals is nonempty, the amplitude vectors are orthogonalized using Cholesky decomposition of the metric $\cramped{S_{\mu\nu}=\langle  \tilde{\Psi}_0| {Y}_\mu {X}_\nu |  \Psi_0 \rangle=\sum_{\rho} L_{\mu\rho} L_{\nu\rho}}$.\cite{doi:10.1063/1.455556,EvangelistaGauss2011,doi:10.1063/1.5040587,fink93} The amplitude update is then given by
$\cramped{\Delta t_\mu = -\sum_{\nu} L_{\mu\nu}^{-1} {\Omega_\nu(t,d,c)}/{\Delta_\nu}}$, where $\Delta_\nu$ is the MP$n$ energy denominator.  The $\lambda$-amplitudes are updated similarly.
The CI-expansion vectors are updated with aid of a minimum polynomial extrapolation technique.\cite{doi:10.1137/0713060} In the bivar-MRECC case, a flexible microiteration scheme is employed.
\section{Numerical Results}
\label{sec:numerics}

Accuracy is one of the most important requirements for a novel MRCC method. Other (weaker) conditions have also been formulated.\cite{Lyakh2012}  
Moreover, being partially motivated from \emph{mathematical} arguments and based on an unconventional formulation of quantum mechanics, the bivar-MRCC method should be tested with respect to \emph{physical predictions}, i.e., expectation values as defined in Eq.~\eqref{eq:bivp-expt-val}.
For this reason, we did not only investigate the accuracy of absolute energies, but also the quality of the actual density operators $\rho = \ket{\Psi}\bra{\tilde{\Psi}}/\braket{\tilde{\Psi}|\Psi}$ compared to FCI results. Furthermore we emphasize, that our scope is not to present the performance of the method under ideal conditions, but rather study the behaviour using set-ups typically found in everyday and sub-optimal applications, e.g., by using different reference orbitals. To this end, absolute energies, spin-expectation values, dipole moments and density operators have been computed using different orbital sets and compared to FCI results. Whenever possible, computed quantities are compared to the results of other MRCC methods found in the literature.

\subsection{Error measures relative to full CI}
\label{sec:error-measures-numerics}
Absolute energies are compared to the respective FCI values by evaluating  the difference $\cramped{\Delta E_\text{FCI} = \text{Tr}(\rho H) - E_\text{FCI}}$. 
In order to quantify the accuracy of the density operator, the Frobenius norm
$\cramped{|| \delta \rho_\text{FCI} ||_\text{F}^2 = { \text{Tr}\bigl((\rho - \rho_\text{FCI})^\dagger  (\rho - \rho_\text{FCI})\bigr)}}$ has been calculated, a standard coherence and entanglement measurement in quantum information theory.\cite{2007291frob,frob2016} Small values of $ || \delta\rho_\text{FCI} ||_\text{F}$ indicate that $\rho$ is a good approximation to the FCI state. Since spin-symmetry can be directly related to the quality of the approximate bra and ket,\cite{doi:10.1063/1.1308557} the total-spin contamination $\cramped{\Delta {S^2}_\text{FCI} = \bigl(  \text{Tr}(\rho S^2) -S(S+1)\bigr) /\hbar^{2}}$
is used as an additional accuracy measure. Finally, the accuracy of dipole moments is expressed by $\cramped{|| \delta m_\text{FCI}||_2^2 =\sum_{i=1}^3 (\text{Tr}(\rho m_i) - m_{i,\text{FCI}})^2/(e a_0)^2}$, where $m_i$ denotes the $i$-th component of the electronic dipole operator. 

 Statistical errors are computed using the following definitions: deviation $\cramped{\Delta x_i = x_i-x_i^\text{FCI}}$, mean deviation, $\cramped{\overline{\Delta x_i} = \sum_{i=1}^n \Delta x_i/n}$, mean absolute deviation $\cramped{\text{MAD}(\Delta x_i) =  \sum_{i=1}^n |\Delta x_i|/n}$, maximum absolute deviation $\cramped{\text{MAX}(\Delta x_i) = \max_{i=1,\dots,n} |\Delta x_i|}$, standard deviation $\cramped{\text{STD}^2(\Delta x_i) = { \sum_{i=1}^n (\Delta x_i - \overline{\Delta x_i})^2}/(n-1)}$ and non-parallelity error $\cramped{\text{NPE}(\Delta x_i) = \max_{i=1,\dots,n} \Delta x_i - \min_{i=1,\dots,n} \Delta x_i}$.

\subsection{Model systems}

A multireference method should be be reasonably accurate for both singlereference and multireference problems.  We therefore opted for studying the novel bivar-MR(E)CC methods using a model system providing both. The computational investigation of the potential curve of the symmetrical insertion of Be into H$_2$ has been described comprehensively\cite{purvis83} and serves as a standard example for testing novel MRCC methods, since it is computationally feasible even for complicated methods.\cite{evangelista11,Lyakh2012} Moreover, it features a lot of problems for quantum chemical methods like severe multireference character, level crossings and change of leading determinant along the potential curve.\cite{Koehn2012} Therefore the performance of the bivar-MR(E)CC methods has mainly been tested using this system. Additionally, the chemical bond-breaking of the hydrogen flouride molecule in ground state and the widely-used H$_8$ model system\cite{Jankowski1985,doi:10.1063/1.481649} have been studied using the bivar-MRCC(2,2)FOIS method yielding highly accurate results. For example, H--F bond breaking with 12 points in $1.0 \le R_\text{H--F} \le 5.0~a_0$ yielded $\text{MAD}(\Delta E_\text{FCI} )= 1.12~mE_\text{H}$, $\text{MAX}(\Delta E_\text{FCI} )= 1.27~mE_\text{H}$, $\text{STD}(\Delta E_\text{FCI} )= 0.06~mE_\text{H}$, and $\text{NPE}(\Delta E_\text{FCI} )= 0.22~mE_\text{H}$ (cf. Section~II, SI). However, since the electronic structures of HF and H$_8$ are less complicated then the one of the BeH$_2$ system, the results are not discussed in detail here, but can be found in the Supplementary Information. 

\subsection{Technical details}

In order to be able to compare to other MRCC methods and owing to computational restrictions, the same parameters regarding geometry and basis set described in Refs.~\onlinecite{evangelista11,adamowicz15,Jankowski1985} have been used for the 
BeH$_2$ (10s3p/3s2p and 4s/2s basis), HF (DZV basis) and H$_8$ (minimal basis) models. The ($C_1$ as well as $C_{2v}$) CASSCF orbitals have been computed using the Bochum-suite of \emph{ab initio} wave function programs.\cite{bochum2,bochum1,fink93} The FCI computations are performed with a local program based on a quasi-relativistic CI program\cite{phdtilmann} and verified against the results presented in Ref. \onlinecite{evangelista11}.
 If not otherwise mentioned, energies and residuals were converged to  thresholds $\cramped{10^{-6}}~\text{a.u.}$ and $\cramped{10^{-4}}~\text{a.u.}$, respectively. Amplitudes with absolute value smaller than $\cramped{10^{-10}}~\text{a.u.}$ were neglected.  In all computations, all electrons were correlated.

\subsection{BeH$_2$ FCI results}\label{sec:num-beh2-fci}

\begin{figure}[htb]
\centering
\includegraphics[width=0.45\textwidth]{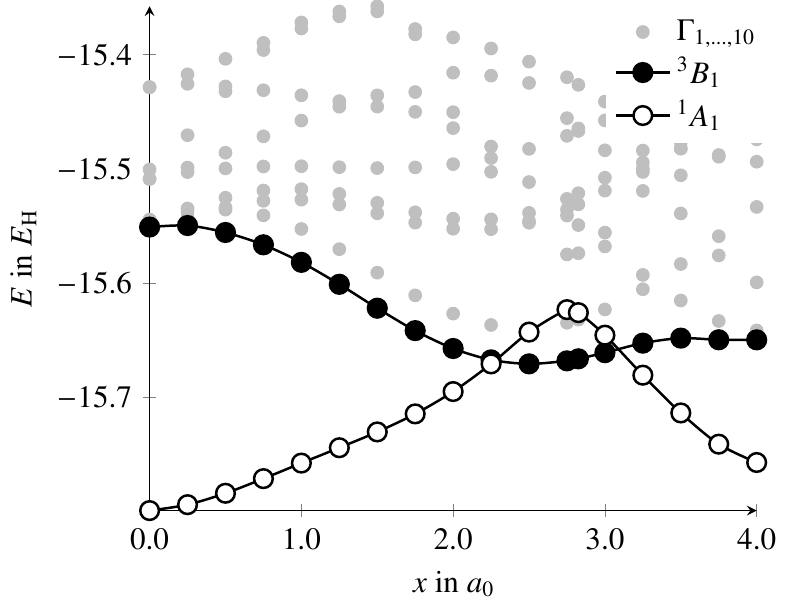}
\caption{FCI potential curves for the $C_{2v}$-symmetric insertion of Be into H$_2$. Note the region where the $\cramped{^1A_1}$ state becomes an excited state.}
\label{fig:fcipes}
\end{figure}

The symmetric insertion geometry of the system is parameterized using the distance $x$ from the Be atom to the H$_2$ moiety, with $x=0$ referring to the linear arrangement.\cite{evangelista11} (Note that the molecule has been rotated in space, thereby interchanging $b_1$ and $b_2$ irreducible representations.)
For $0 <x\le4~a_0$, the system comprises the symmetry of the $C_{2v}$ point-group. The FCI energies for the first 10 states with $M_S=0$ are shown in Fig.~\ref{fig:fcipes}. (The  mapping of the $C_1$ FCI states  $\Gamma_1, \Gamma_2$, etc., onto the corresponding states in $C_{2v}$ is given in the SI.) Apparently, the nature of the $C_1$ ground state changes along the insertion pathway. The respective $C_1$ ground state $\Gamma_1$ ($\cramped{^1A_1}$ or $\cramped{^3B_1}$ in $C_{2v}$) is dominated by the appropriate linear combinations of the  following four determinants:
\begin{equation}\label{eq:fcidets}
\begin{split}
    |\Phi_1\rangle &= |(1a_1)^2 (2a_1)^2 (1b_1)^2 \rangle, \\
    |\Phi_2\rangle &= |(1a_1)^2 (2a_1)^2 (3a_1)^2 \rangle,  \\
        |\Phi_{3,4}\rangle &= |(1a_1)^2 (2a_1)^2 (3a_1)^1(1b_1)^1 \rangle,
    \end{split}
\end{equation}
where the exponent denotes the (spatial) orbital occupancy.
Concerning  the cusp by $x=2.75~a_0$, the FCI wave function of the $\cramped{^1A_1}$ state (CAS(2,2)SCF orbitals) constitutes ${\sim}52\%$ of $|\Phi_1\rangle$ and ${\sim}39\%$ of $|\Phi_2\rangle$, making it more suitable for a singlereference based MRCC description than a 50:50 mixture. Furthermore, there are small contributions from other determinants, namely the $^1A_1$-symmetric combinations of $\cramped{|(1a_1)^2(2a_1)^1 (3a_1)^1 (1b_1)^1 (2b_1)^1\rangle}$ (${\sim}2\%$) and $\cramped{|(1a_1)^2(2a_1)^2  (1b_2)^2 \rangle}$ (${\sim}2\%$) which should be included in an accurate correlation treatment (cf. Section~III, SI).

\subsection{Absolute energies}

The potential energy curve of the  $\cramped{^1A_1}$ state has been investigated using the bivar-MRCCSD method with 2-in-2 (CAS(2,2)) and 4-in-6 active spaces (CAS(4,6)), based on the respective CASSCF orbitals.  The CAS(2,2) is spanned by the four determinants given in Eq.~\eqref{eq:fcidets}, i.e., the $1b_1$ and $3a_1$ orbitals are chosen active. For CAS(4,6), the doubly occupied $2a_1$ orbital and three virtual  orbitals are added based on orbital energy\cite{Roosbook}.
The reference wave functions for the bivar-MR(E)CCSD computations using the CAS(2,2) model space have been constructed in the following way (The same procedure has also been applied for the computations based on the CAS(4,6) model space): Diagonalizing both the Hamiltonian and the $K$-matrix in this space yields four states, that in $C_{2v}$ transform as two  $\cramped{^1A_1}$, one $\cramped{^1B_1}$, and one $\cramped{^3B_1}$ state. The initial values for the expansion vectors $c$ and $d$ were chosen such that they correspond to the one of the two the $\cramped{^1A_1}$ (CASCI) states being lowest in energy. The reference wave functions were then updated during the iterative procedure,  preserving the $\cramped{^1A_1}$ nature of the expansion vectors throughout by an overlap criterion (cf. Section ~\ref{sec:reference-wave-function-during-iteration}).
  Note, that for $2.25\le x \le 3.0~a_0$, this means that the optimization has been performed for an excited state.

\begin{figure}[htb]
\centering
\includegraphics[width=0.45\textwidth]{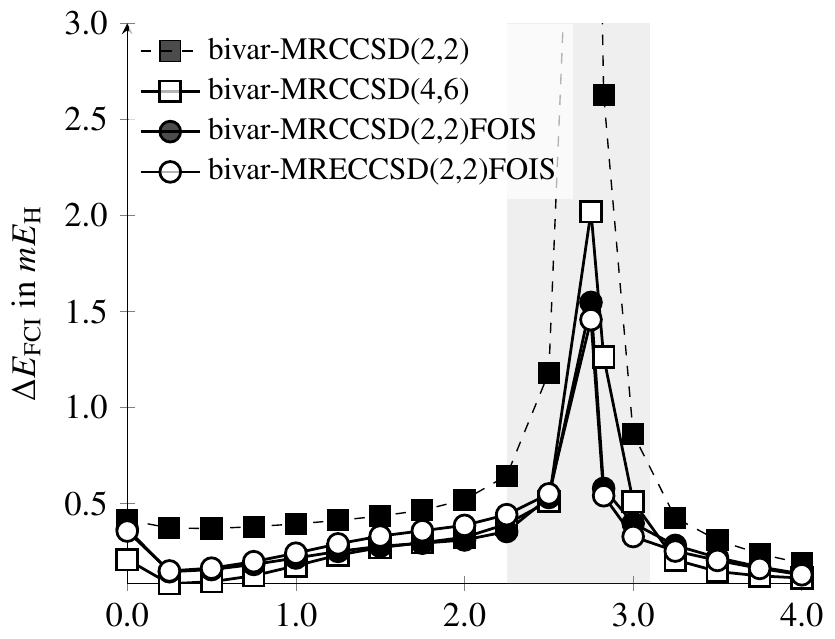}
\caption{bivar-MR(E)CCSD energy differences with respect to FCI for the $\cramped{^1A_1}$ state of BeH$_2$ based on the respective CASSCF orbitals. The light gray background indicates the region where  $\cramped{^1A_1}$ is an excited state.
\label{fig:bivarmrccsdresultsa1}}
\end{figure}

The results can be found in Fig.~\ref{fig:bivarmrccsdresultsa1}.  The bivar-MRCCSD(2,2) results are close to the values obtained with singlereference CCSD using $|\Phi_1\rangle$,\cite{evangelista11} indicating that these calculations lack important dynamical correlation contributions from other determinants, in particular doubly excited determinants relative to  $|\Phi_2\rangle$.\cite{doi:10.1063/1.481649} Including single- and double excitations for all four model space determinants (bivar-MRCCSD(2,2)FOIS) improves the treatment significantly and decreases the maximum error to just under the ``chemical accuracy limit'' of $1~\text{kcal}/\text{mol}$ (${\sim}1.594~mE_\text{H}$). 
The results of the extended variant  bivar-MRECCSD(2,2)FOIS are very similar (cf. Table~\ref{tab:falspos}), slightly superior in the multireference, but slightly inferior in the singlereference region. The same findings have been described for the singlereference coupled cluster and extended coupled cluster methods using the same model system.\cite{evangelista11} Increasing the model space to include 225 determinants (CAS(4,6)), but neglecting the FOIS (bivar-MRCCSD(4,6)) yields results similar to the four determinant model space variant with additional FOIS (bivar-MRCCSD(2,2)FOIS).

All curves show a discontinuity at $x=2.75~a_0$ which can be traced back to the complicated nature of the FCI wave function at this geometry, constituting an almost 50:50 mixture of the determinants $|\Phi_1\rangle$ and $|\Phi_2\rangle$ as described in Section~\ref{sec:num-beh2-fci}. 
The effect of bivariational reference optimization by orbital optimization at this point will be discussed in Section~\ref{sec:orbitaloptimization}. 

\begin{figure}[htb]
\centering
\includegraphics[width=0.45\textwidth]{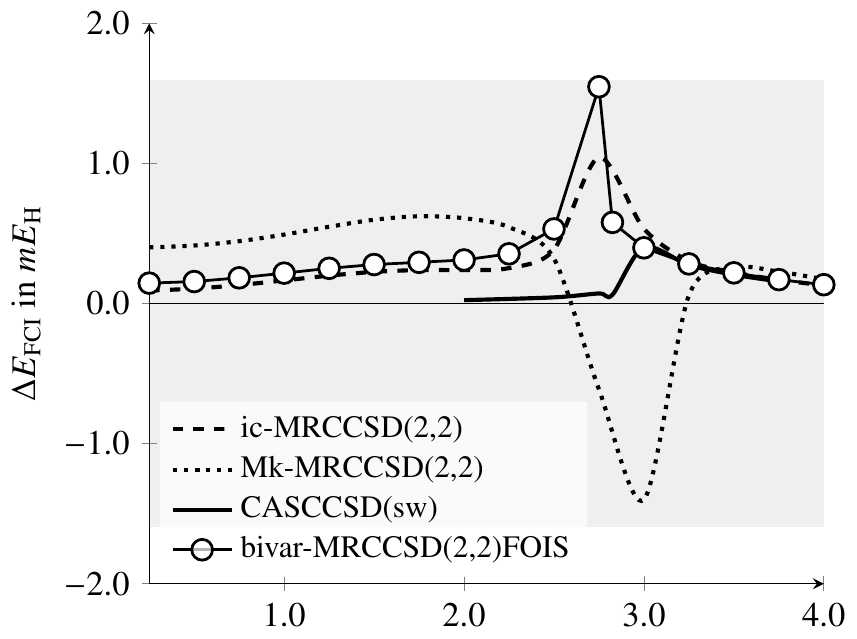}
\caption{Method comparison showing energy differences of the $\cramped{^1A_1}$ state of BeH$_2$ with respect to FCI results of different MRCC methods. The gray area indicates the region of "chemical accuracy", i.e., $|\Delta E_\text{FCI}| \le 1~\text{kcal}/\text{mol}$.
\label{fig:methodcomparison}}
\end{figure}

A comparison to other MRCC methods using identical basis/geometry setup is shown in Fig.~\ref{fig:methodcomparison}.  The ic-MRCCSD method uses a sophisticated internally contracted ansatz  where the cluster operator includes terms from all model space states, see Eq.~\eqref{eq:ic-mrcc}.\cite{Hanauer2011,EvangelistaGauss2011} It can therefore be assumed to be similar or more accurate than the  singlereference ansatz used in bivar-MRCCSD. The latest CASCCSD(sw) method is closest to our current bivar-MRCCSD(2,2)FOIS method but much more accurate in the region where $\cramped{^1A_1}$ is an excited state in $C_1$.\cite{adamowicz15} This discrepancy might primarily be traced back to the different model space reference wave function used. In test computations on a H$_8$ model system, the CASCCSD(sw) and MRCCSD(2,2)FOIS results where very similar, with a slight superiority of MRCCSD(2,2)FOIS in the multireference region (cf. Section~I, SI). Additionally, the results using the established Mk-MRCCSD method\cite{Mahapatra1999} are shown. However, being an Jeziorski--Monkhorst type method\cite{Jeziorski1981}, a direct comparison is complicated, and we merely note that the overall accuracy is good despite the instability in the region where $\cramped{^1A_1}$ is not the ground state. Finally, we also like to mention that the MRCCSD method from K\'{a}llay, Szalay, and Surj\'{a}n\cite{Kallay2002} ($\Delta E_\text{FCI} = 1.890~m E_\text{H}$ at $x=2.75~a_0$) and the 
MRexpT method from Hanrath \emph{et al.}\cite{Hanrath2005} (MAD($\Delta E_\text{FCI}) = 0.591~m E_\text{H}$, MAX($\Delta E_\text{FCI}) = 1.693~m E_\text{H}$, NPE($\Delta E_\text{FCI}) = 1.663~m E_\text{H}$) are very accurate. However, the reported values are based on SCF orbitals and/or use a different basis set and are therefore not shown here. Altogether, all multireference MRCCSD methods discussed, including the novel bivar-MRCCSD model, demonstrate chemical accuracy, i.e., $|\Delta E_\text{FCI}|\le 1~\text{kcal}/\text{mol}$ for this model system.

\subsection{Density operators}

% \tbcomment{OLD:In particular when the desired state is not the ground state, absolute energies are a poor measurement for accuracy. In Fig.~XY in SI, the energy differences of several bivar-MRCCSD(2,2)FOIS computations with respect to FCI values are shown. These computations differ only in the CAS orbitals  Apparently, using the ``right'' set of orbitals, one can get arbitrarily close to and even below the FCI value in the region where the desired state is not the ground state. Thus, in order to asses, if get the right value is obtained for the right reason, a more reliable characteristic has to be used, e.g., the Frobenius norm of the difference density operator, $||\delta\rho_\text{FCI}||_\text{F}$ as described in Section~\ref{sec:error-measures-numerics}.}

\begin{figure}[htb]
\centering
\includegraphics[width=0.45\textwidth]{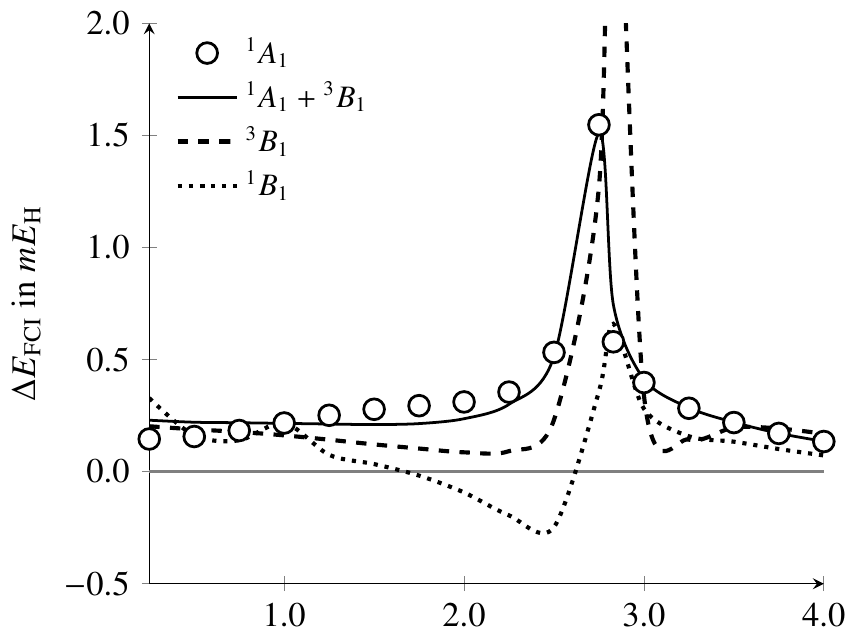}
\caption{bivar-MRCCSD(2,2)FOIS energy differences with respect to FCI for the $\cramped{^1A_1}$ state of BeH$_2$ based on CAS(2,2)SCF orbitals optimized for different electronic states.}% The light gray background indicates the region where  $\cramped{^1A_1}$ is an excited state.}%} The dotted line denotes "chemical accuracy".}
\label{fig:num-22fois-orbcomp}
\end{figure}

Absolute energies are not good indicators of accuracy in general, particularly when the desired state is not the ground state. In Fig.~\ref{fig:num-22fois-orbcomp}, the energy differences of several bivar-MRCCSD(2,2)FOIS computations with respect to FCI values are shown. These computations differ only in the orbitals used for constructing the computational Hilbert space, including the external space.  Apparently, by using the ``right'' set of orbitals one can get very close to and even below the FCI energy, in particular inside the region where the desired state is not the ground state. Thus, in order to asses whether the right value has been obtained for the right reason, a more reliable characteristic has to be used.

To this end, the Frobenius norm $||\delta\rho_\text{FCI}||_\text{F}$ of the difference density operator (cf.~Section~\ref{sec:error-measures-numerics}) of the $\cramped{^1A_1}$ state has been computed for the bivar-MRCCSD(2,2)FOIS method along the BeH$_2$ potential curve using CAS(2,2)SCF orbitals optimized for different electronic states. This data is summarized in Tab.~\ref{tab:falspos}. Additionally, values for the bivar-MRECCSD(2,2)FOIS and bivar-MRCCSD(4,6) variants are shown for comparison.

The mean absolute deviations of both the bivar-MRCCSD(2,2)FOIS energy and density errors computed over the entire potential curve using different orbitals are similar, but the maximal absolute deviations differ significantly. Consider for example the errors at $x=2.75~a_0$. While the energy error of the computation based on orbitals for $\cramped{^1B_1}$ is very small ($0.358~mE_\text{H}$), the error in the density operator is rather large ($0.206$) when compared to the errors obtained with $\cramped{^1A_1}$ orbitals ($1.547~mE_\text{H}$ and $0.052$).  This can be resolved by analysing the CAS(2,2)SCF wave functions: %The active space is generated by the determinants $|\Phi_1\rangle \dots |\Phi_4\rangle$ that can be constructed by distributing two electrons in the two orbitals $3a_1$ and $1b_1$ (cf. Section~\ref{sec:num-beh2-fci}).
 The $\cramped{^1B_1}$ state is composed of the open-shell determinants $|\Phi_3\rangle$ and $|\Phi_4\rangle$ (cf. Eq.~\eqref{eq:fcidets}), while in $\cramped{^1A_1}$, the weights of the closed-shell determinants $|\Phi_1\rangle$ and $|\Phi_2\rangle$ are large. Thus, optimizing orbitals for the $\cramped{^1B_1}$ (or $\cramped{^3B_1}$) state has a significant impact on the  $\cramped{^1A_1}$ wave function without contaminating the overall symmetry of the wave function.
 
 Based on the density error analysis, it can thus be concluded that the energy errors of the  computations with orbitals optimized for $\cramped{^1B_1}$ and $\cramped{^3B_1}$ are unreliable.  In contrast, the values obtained with state-specific ($\cramped{^1A_1}$) and state-average (50:50 mixture of $\cramped{^1A_1}$ and  $\cramped{^3B_1}$ states) orbitals are considerably smaller,  i.e., they represent the FCI state better.
Considering actual applications, we note that typically one of the latter two orbital sets will be used\cite{Roosbook} --  both of which have been demonstrated to be accurate for the right reason. 

\begin{table}[htb]\centering
\caption{Comparison of energy and density operator  error characteristics of the BeH$_2$ potential curve computed with bivar-MRCCSD(2,2)FOIS based on CAS(2,2)SCF orbitals optimized for different states. For comparison, bivar-MRECCSD(2,2)FOIS and bivar-MRCCSD(4,6) values are presented in the last two rows.}
\begin{ruledtabular}
\begin{tabular}{crrrrrr}
Orbitals & \multicolumn{3}{c}{$\Delta E_\text{FCI}$ ($mE_\text{H}$)} & \multicolumn{3}{c}{$|| \delta \rho_\text{FCI} ||_\text{F}$} \\
~& MAD & MAX & $2.75~a_0$ & MAD & MAX & $2.75~a_0$  \\\hline
$\cramped{^1A_1}$ & $0.356$ & $1.547$ & $1.547$ &$0.020$ & $0.052$ & $0.052$ \\ 
{$\cramped{^1A_1 + {^3B_1}}$\footnote{50:50 mixture}} & $0.356$ & $1.511$ & $1.511$ & $0.021$ & $0.040$ &  $0.033$\\ 
$\cramped{^3B_1}$ & $0.411$ & $3.172$ & $1.272$ & $0.033$ &$0.199$  &  $0.108$\\ 
$\cramped{^1B_1}$ & $0.192$ & $0.659$ & $0.358$ & $0.037$ &$0.206$  &     $0.206$\\\hline
{$\cramped{^1A_1}$\footnote{bivar-MRECCSD(2,2)FOIS}}& $0.364$ & $1.458$ & $1.458$ & $0.021$ & $0.050$ & $0.050$ \\ 
{$\cramped{^1A_1}$\footnote{bivar-MRCCSD(4,6)}} &   $0.406$ & $2.020$ & $2.020$ & $0.023$ & $0.078$ &  $0.078$\\ 
\end{tabular}
\end{ruledtabular}
\label{tab:falspos}
\end{table}
 
As a second measure for density operator accuracy, the  spin-contamination $\cramped{\Delta {S^2}_\text{FCI}}$ has be computed. In the present implementation, only the symmetry $[H,S_z]=0$ is exploited, total-spin conservation is not enforced.
 However, using a qualitatively correct model-space bra and ket is generally thought to be able to reduce the spin-contamination in the correlated wave function substantially.\cite{Lyakh2012} For all singlet states studied, the computed spin-contamination was negligible with  $\cramped{\Delta {S^2}_\text{FCI}\ll 10^{-3}}$. Concerning the triplet $\cramped{^3B_1}$ state, the errors are relatively small for bivar-MRCCSD(2,2)FOIS computations ($\text{MAD}(\cramped{\Delta {S^2}_\text{FCI}}) < 0.02$, $\text{MAX}(\cramped{\Delta {S^2}_\text{FCI}}) < 0.07$, cf. Section~III, SI) compared to the values discussed in the context of singlereference CC methods.\cite{doi:10.1063/1.467312,doi:10.1063/1.1308557,doi:10.1063/1.468144} These errors decrease further with increasing active space size ($\text{MAD}(\cramped{\Delta {S^2}_\text{FCI}}) < 0.01$, $\text{MAX}(\cramped{\Delta {S^2}_\text{FCI}}) < 0.04$ for bivar-MRCCSD(4,6)). Thus, the errors in the expectation value of the total spin-operator induced by the coupled cluster expansion used in the bivar-MRCC methods are, at least for this case, insignificant.

\subsection{Orbital optimization}\label{sec:orbitaloptimization}

\begin{figure}[htb]
\centering
\includegraphics[width=0.45\textwidth]{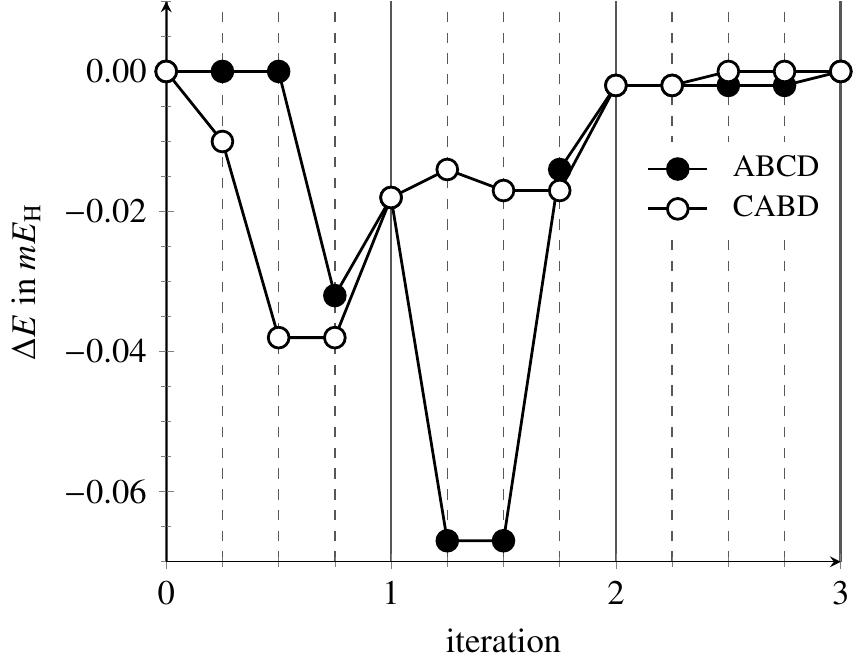}
\caption{Deviation of two different bivar-OAMRCCSD(4,6) from a bivar-MRCCSD(4,6) computation for the BeH$_2$ model at $x=2.75~a_0$ in $mE_\text{H}$ ($\cramped{^1A_1}$ state, CAS(4,6)SCF orbitals). Steps per iteration: A ($t$-equations), B ($\lambda$-equations), C (orbital optimization), D (non-symmetric eigenvalue problem).%s
}
\label{fig:orb-opt}
\end{figure}

In the orbital-adaptive variant bivar-OAMRCC, active-active non-orthogonal orbital rotations are introduced via Eq.~\eqref{eq:bivar-MRCC-working-orb-transform}. Since these transformations are restricted to the model space, one may argue that these compete with the diagonalisation of the $K$-matrix Eq.~\eqref{eq:bivar-MRCC-working-evp}. In order to investigate this, test computations have been performed on the BeH$_2$ system in the multireference region ($x=2.75~a_0$) using the bivar-OAMRCCSD(2,2)FOIS and bivar-OAMRCCSD(4,6) models. In all calculations the results have been found to be very close to the results obtained without orbital optimization.
For the computations using the small CAS(2,2) model space, this is due to almost vanishing gradient norms. Concerning the CAS(4,6) based bivar-OAMRCCSD(4,6) computations, the results can be analysed by considering the energy change during iterations. To this end, each iteration is divided into four parts, namely solving $t$-equations (A), $\lambda$-equations (B), non-orthogonal orbital optimization (C) and non-symmetric eigenvalue problem (D). At each step, the energy difference between two bivar-OAMRCCSD(4,6) and bivar-MRCCSD(4,6) computations for the $\cramped{^1A_1}$ state is shown in Fig.~\ref{fig:orb-opt} (values for the $\cramped{^3B_1}$ state can be found in the SI). The two orbital-adaptive models differ in the partition ordering,  i.e., whether the orbital optimization is conducted before (CABD model) or after solving the CC equations (ABCD model).

Indeed it seems like orbital optimization and diagonalization compete in the first iterations, i.e., orbital optimization leads, for both models, to a slight energy lowering of the energy, whereas the diagonalization has the opposite effect. In the second iteration of the computation using ABCD ordering, also the orbital optimization  leads to an energy increase. In total, the effect of orbital optimization is very small after the third optimization, i.e., the bivar-OAMRCCSD(4,6) results are almost identical to the ones obtained without orbital optimization. Thus, in the example investigated, orbital adaptivity has negligible effect. However, this insensitivity of the energy towards bivariational optimization of the reference may be due to its limited size. The SD-FOIS truncation of the external amplitudes is already very large for a six-electron system. Recall, that at the FCI limit all orbital rotations are redundant. Moreover, the choice of the reference should play a major role in general, since the conventional argument against MRCC methods based on single-reference theory is its bias towards the formal reference. Thus, we conjecture that for larger systems, in particular extended systems, orbital optimization will play a much larger role.

\subsection{Molecular properties}

\begin{figure}[htb]
\centering
\includegraphics[width=0.45\textwidth]{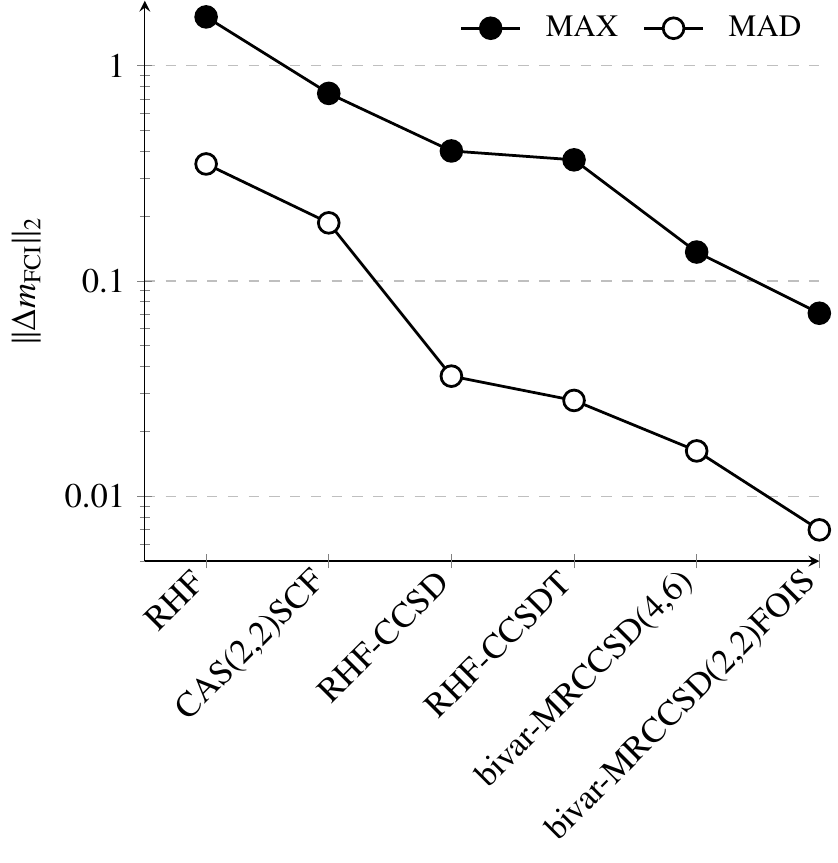}
\caption{Error characteristics of the electronic dipole moment vector compared to FCI values for the potential curve of the $\cramped{^1A_1}$ state of BeH$_2$.}
\label{Fig:diperr}
\end{figure}

To gain some insight into the accuracy of first-order properties computed with the bivar-MRCC method, 
the  dipole moments of the $\cramped{^1A_1}$ state of BeH$_2$ along the PES have been computed using Eq.~\eqref{eq:bivp-expt-val}. The electronic dipole moment integrals were taken from a local version of the TURBOMOLE program package.\cite{doi:10.1002/jcc.21692} The computed values are compared to the corresponding FCI results by evaluating the error measure $||\Delta m_\text{FCI}||_2$ (cf. Section~\ref{sec:error-measures-numerics}). For comparison, (orbitally unrelaxed) singlereference CC dipole moments have been computed using the CFOUR program package.\cite{cfour} The mean absolute and maximum absolute deviations are depicted in Fig.~\ref{Fig:diperr}, the individual values can be found in the SI. The CCSD and CCSDT dipole moments based on restricted Hartree-Fock orbitals are very accurate for singlereference systems, but less accurate in the multireference region.\cite{leescus} The bivar-MRCCSD computations improve upon the CAS(2,2)SCF reference values significantly, and even outperform the RHF-CCSDT method in this example.

\section{Conclusion}
\label{sec:conclusion}

In this article, we have introduced a state-specific multireference coupled-cluster method based on Arponen's bivariational principle, the bivar-MRCC method. An extended version, bivar-MRECC, was also discussed, similar to Arponen's extended CC method. The method is wholly based on singlereference theory, has modest complexity, and avoids some of the problems associated with established multireference methods. For example, all cluster operators commute, and there is no need for sufficiency conditions as is needed in, e.g., Mukherjee's state-specific method. The method requires a formal reference much like the CASCC method of Piecuch, Oliphant, and Adamowicz, but the bias is to a great extent eliminated using bivariational optimization of the reference.

An important aspect of the method is its manifest separability, which in the extended form is exact, a feature allowed by the bivariational approach only. We expect the separability to play a major role for excited states and response properties. 

A pilot implementation has been described, and extensive benchmark calculations on the insertion of a Be atom into H$_2$ has been performed. Therein, the method has been demonstrated to be very accurate, i.e., within the desired range of chemical accuracy,  and an analysis of the density-  and total-spin operator shows that the state description is indeed very accurate despite that the ansatz requires ``two wave functions''. All in all, the bivar-MRCC method seems to perform equally well as established state-specific MRCC methods do.
While the pilot implementation is based on full-configuration interaction methodology to facilitate rapid development of a flexible program, a more efficient and optimally scaling implementation has been outlined. Such an implementation will open up the possibility for applications closer to the state-of-the-art, including transition metal chemistry and luminescence phenomena.

The systems studied exhibited only weak dependence on the orbital rotations in the working equations. We conjecture that for such small systems as were studied, the first-order interaction space is sufficient to describe the majority of dynamical correlation, which means that the bivar-MRCC state is near the FCI state, and in this limit the orbital rotations are redundant. We furthermore conjecture that orbital rotations will play a larger role for larger systems.

From the point of view of theory, the natural continuation of this work is the derivation of response theory and theory for excited states, which the bivariational approach allows in a relatively straightforward manner. Moreover, the time-dependent bivariational principle combined with biorthogonal orbital-optimization allows an \emph{ab initio} dynamics method suitable to, say, study molecules under the influence of intense laser pulses, charge migration, charge transfer, and other situations where the system evolves far away from the ground-state. Multireference character of the resulting state can be expected to be significant, and is a major challenge of state-of-the-art methods today.

The bivariational formulation of bivar-MRCC has an important advantage in that a mathematical \emph{a priori} error analysis is possible.  The major challenge is finding the right assumptions on the system Hamiltonian and model space to facilitate a monotonicity analysis. If these assumptions are also reasonable in a wide range of situations, the bivar-MRCC method gains a distinct advantage over other MRCC theories, for which few mathematical results exist.

The modest complexity of the bivar-MRCC method allows extending the field of application far beyond the simple benchmark calculations presented here, once an efficient implementation is in place. We conclude that the bivar-MRCC method has potential to become a useful and practical tool in many areas of quantum molecular sciences, also for non-experts.

\begin{acknowledgments}
This work has received funding from the Research Council of Norway (RCN) under
CoE Grant Nos. 287906 and 262695 (Hylleraas Centre for Quantum Molecular Sciences), and from ERC-STG-2014 under grant agreement No 639508. We further thank the Norwegian Metacenter for Computational Science for support via NOTUR, project no. NN4654K. The authors thank A.~Laestadius, M.~E.~Harding and T.~B.~Pedersen for constructive remarks.
\end{acknowledgments}

\section*{Data Availability}
The data that supports the findings of this study are available within the article and its supplementary material.

\bibliography{bivaqum-mrcc}
\bibliographystyle{aipnum4-1}

% % https://tex.stackexchange.com/questions/227385/inserting-a-pdf-file-in-revtex-4-1-with-reprint-option


\section*{Supplementary material}

**Supplementary information: A state-specifc multi-reference coupled-cluster method based on the bivariational principle**

T. Bodenstein[1] and S. Kvaal[1, a)]

*Hylleraas Center for Quantum Molecular Sciences, Department of Chemistry, University of Oslo, Norway*

---

a)Electronic mail: simen.kvaal@kjemi.uio.no.

# I. H$_8$ MODEL SYSTEM

In order to compare the new MRCC method based on the bivariational principle with the CASCC method from Adamowicz *et al.*, we implemented the disconnected CASCCSD equations as described in (L. Adamowicz, J.-P. Malrieu, V. V. Ivanov, *J. Chem. Phys.* 112, 10075 (2000)):

$$0 = \langle\Phi_0|Y_\mu^{\text{int}}(H - \Delta E_{\text{CASCC}})e^T(1+C)|\Phi_0\rangle$$

$$0 = \langle\Phi_0|Y_\mu^{\text{ext}}(H - \Delta E_{\text{CASCC}})e^T(1+C)|\Phi_0\rangle$$

$$\Delta E_{\text{CASCC}} = \langle\Phi_0|He^T(1+C)|\Phi_0\rangle$$

in our program. Here, $C$ contains single and double excitations within the model space, whereas $T$ is identical to the external cluster operator used in bivar-MRCCSD including contributions from the FOIS. Consequently, $Y^{\text{int}}$ is a pure internal de-excitation operator where as $Y^{\text{ext}}$ contains at least one inactive or external index. The equations were solved iteratively starting from MP2 values for the $t_2$-amplitudes and CASSCF values for the CI-coefficients. For technical reasons, we validated our implementation against reported CASCCD results (V. V. Ivanov, L. Adamowicz, *J. Chem. Phys.* 113, 8503 (2000)) on the H$_8$ model system (K. Jankowski, L. Meissner, J. Wasilewski, *Int. J. Quantum Chem.* 28, 931 (1985)). This system is widely used to study the performance of multireference methods because it depends on a geometrical parameter $\alpha$ which can be used to tune the multireference character of the system where $\alpha = 0$ is essential singlereference and $\alpha = 0.0001$ and smaller generates high multireference character. We compared the CASCCSD method with the new bivar-MRCCSD(2,2)FOIS method using these two values, converging all equations iteratively in a sequence of CI-, and CC-microiterations followed by an energy update to a energy difference and residual threshold of $10^{-8}$ a.u., and $10^{-6}$ a.u., respectively using the same reference determinants and identical integrals. Expansion parameters were neglected if their absolute value was smaller than $10^{-16}$. The results are summarized in the tables below. In the singlereference case, both methods are, as expected, very similar. For the multi-reference case however, the bivar-MRCCSD(2,2)FOIS is slightly better with respect to both, energy and state projector compared to CASCCSD.

TABLE I: Results for the H$_8$ system at $\alpha = 1.0$ (singlereference case). Energies given in $E_\text{H}$. The literature values are taken from V. V. Ivanov, L. Adamowicz, *J. Chem. Phys.* 113, 8503 (2000).

| Model | $E$ from Lit. | $E$ this work | $\|\delta\rho_\text{FCI}\|_\text{F}$ |
|---|---|---|---|
| HF | −4.242845 | −4.242846 | 0.484 |
| CAS(2,2)SCF | −4.248955 | −4.248953 | 0.465 |
| CASCCD (w/o FOIS) | −4.352364 | −4.352361 | 0.032 |
| CASCCSD (w/ FOIS) |  | −4.352472 | 0.026 |
| bivar-MRCCSD(2,2)FOIS |  | −4.352473 | 0.027 |
| FCI | −4.352990 | −4.352990 | 0.000 |

TABLE II: Results for the H$_8$ system at $\alpha = 0.0001$ (multireference case). Energies given in $E_\text{H}$. The literature values are taken from V. V. Ivanov, L. Adamowicz, *J. Chem. Phys.* 113, 8503 (2000).

| Model | $E$ in Ref. 1 | $E$ this work | $\|\delta\rho_\text{FCI}\|_\text{F}$ |
|---|---|---|---|
| HF | −4.065562 | −4.065563 | 1.052 |
| CAS(2,2)SCF | −4.082774 | −4.082774 | 0.471 |
| CASCCD (w/o FOIS) | −4.203138 | −4.203138 | 0.191 |
| CASCCSD (w/ FOIS) |  | −4.203725 | 0.063 |
| bivar-MRCCSD(2,2)FOIS |  | −4.203801 | 0.042 |
| FCI | −4.204802 | −4.204803 | 0.000 |

## II. HYDROGEN FLUORIDE POTENTIAL CURVE

The potential curve following the bond-breaking of the H–F molecule has been investigated using the bivar-MRCCSD(2,2)FOIS method (see Tab. III for results). The (DZV) basis set and geometry parameters (i.e., interatomic distance $R_{\text{H--F}}$) were taken from F. A. Evangelista, *J. Chem. Phys.* 134, 224102 (2011). The reference model states were constructed from CAS(2,2)SCF wavefunctions for the ground state. As can be seen from the statistical data, the results are very accurate compared to the FCI values.

TABLE III: bivar-MRCCSD(2,2)FOIS results for the ground state potential curve of the H–F molecule.

| $R_{\text{H--F}}$ in $a_0$ | $E_{\text{FCI}}$ in $E_H$ | $\Delta E_{\text{FCI}}$ in $mE_H$ | $\|\delta\rho_{\text{FCI}}\|_F$ | $\|\Delta m_{\text{FCI}}\|_2$ |
|---|---|---|---|---|
| 1.00 | −99.555487 | 1.202 | 0.020 | 0.002 |
| 1.25 | −99.972491 | 1.163 | 0.019 | 0.003 |
| 1.50 | −100.113237 | 1.145 | 0.019 | 0.004 |
| 1.75 | −100.146980 | 1.131 | 0.019 | 0.005 |
| 2.00 | −100.138919 | 1.115 | 0.019 | 0.006 |
| 2.25 | −100.115745 | 1.094 | 0.019 | 0.007 |
| 2.50 | −100.088755 | 1.073 | 0.019 | 0.007 |
| 2.75 | −100.062850 | 1.056 | 0.019 | 0.007 |
| 3.00 | −100.040048 | 1.051 | 0.019 | 0.006 |
| 3.50 | −100.006065 | 1.073 | 0.019 | 0.004 |
| 4.00 | −99.986559 | 1.113 | 0.019 | 0.001 |
| 5.00 | −99.973059 | 1.272 | 0.021 | 0.000 |
| MAX | – | 1.272 | 0.021 | 0.007 |
| MAD | – | 1.124 | 0.019 | 0.004 |
| STD | – | 0.065 | 0.001 | 0.002 |
| NPE | – | 0.221 | 0.002 | 0.007 |

## III. BE INSERTION INTO $H_2$

The $C_{2v}$-symmetric intertion of a Be atom into the $H_2$ molecule has been investigated using FCI and bivar-MRCCSD methods. In order to allow comparison to established MRCC methods, the same basis set (10s3p/3s2p for Be and 4s/2s for H) and geometry parameter ($y(x) = 2.54a_0 - 0.46x$, where $x$ denotes the distance between the midpoint of the $H_2$–bond and the Be–atom) have been taken from F. A. Evangelista, *J. Chem. Phys.* 134, 224102 (2011). Note, that for technical reasons, the molecule has been rotated in space, such that the the $B_1$ and $B_2$ irreps are interchanged.

### A. CASSCF results

All computations shown below are based on CAS(2,2)SCF and CAS(4,6)SCF orbitals in the $C_{2v}$ point group, respectively. The orbitals have been optimized for different electronic states in order to compare the impact of the orbitals on the correlated bivar-MRCCSD computations. The respective CASSCF energies are listed in Tab. IV.

TABLE IV: CAS(2,2)SCF and CAS(4,6)SCF energies of the BeH$_2$ system in $E_\mathrm{H}$ based on optimization for different states. For the state-specific wavefunctions, energies are given for the respective states, for the state-average orbitals, only the energies of the $^1A_1$ state is shown.

| $x$ in $a_0$ | $^1A_1$(2,2) | $^1A_1 + {}^3B_1$(2,2) | $^3B_1$(2,2) | $^1B_1$(2,2) | $^1A_1$(4,6) | $^3B_1$(4,6) |
|---|---|---|---|---|---|---|
| 0.250 | −15.743737 | −15.736461 | −15.483032 | −15.456248 | −15.768014 | −15.523842 |
| 0.500 | −15.733710 | −15.726431 | −15.488831 | −15.461269 | −15.757862 | −15.529772 |
| 0.750 | −15.720270 | −15.712951 | −15.499732 | −15.472153 | −15.744758 | −15.540590 |
| 1.000 | −15.705567 | −15.698258 | −15.516107 | −15.489106 | −15.730651 | −15.555758 |
| 1.250 | −15.690632 | −15.683482 | −15.537191 | −15.511139 | −15.716907 | −15.574948 |
| 1.500 | −15.674874 | −15.667946 | −15.560440 | −15.535505 | −15.703301 | −15.595915 |
| 1.750 | −15.656459 | −15.649725 | −15.582351 | −15.558517 | −15.687975 | −15.615492 |
| 2.000 | −15.633239 | −15.626620 | −15.599799 | −15.576862 | −15.668880 | −15.631093 |
| 2.250 | −15.603910 | −15.597417 | −15.610821 | −15.588371 | −15.644813 | −15.641163 |
| 2.500 | −15.569738 | −15.562817 | −15.614637 | −15.591988 | −15.618003 | −15.644904 |
| 2.750 | −15.538710 | −15.531034 | −15.611546 | −15.587522 | −15.597666 | −15.642583 |
| 2.825 | −15.536362 | −15.519087 | −15.609470 | −15.584664 | −15.599722 | −15.640804 |
| 3.000 | −15.558524 | −15.548494 | −15.603209 | −15.575462 | −15.620528 | −15.635453 |
| 3.250 | −15.599783 | −15.592021 | −15.593915 | −15.556841 | −15.655549 | −15.626970 |
| 3.500 | −15.637371 | −15.633329 | −15.591698 | −15.533188 | −15.688039 | −15.622274 |
| 3.750 | −15.668293 | −15.666337 | −15.598094 | −15.506780 | −15.714556 | −15.622865 |
| 4.000 | −15.687220 | −15.686234 | −15.602530 | −15.480540 | −15.729704 | −15.622978 |

## B. FCI results

The FCI energies of the four lowest states are shown in Tab. V. The mapping of the three lowest $C_1$ FCI states with $M_S = 0$ onto the respective irreps in $C_{2v}$ is listed in Tab. VI. Note that in the region $2.25 < x \leq 3.0$, the $^1A_1$ state is not the ground state.

The FCI wave functions were computed based on CAS(2,2)SCF wave functions (see Section. III A). The change of the leading configuration in the $^1A_1$ state is shown in the left side of Fig. 1, the right side shows the composition of the $^3B_1$ state. Note that for symmetry reasons, the weights of the two determinants in $^3B_1$ are identical. The composition of the FCI wavefunction at $x = 2.75$ $a_0$ is indicated in Fig. 2. As can be seen, apart from the two determinants shown in Fig. 1, there are also three other determinants with a significant weight in the $^1A_1$ FCI wavefunction. An accurate dynamical correlation treatment has to include these.

TABLE V: FCI energies of the BeH$_2$ system in $E_H$ ($M_S = 0$).

| $x$ in $a_0$ | FCI state 1 | FCI state 2 | FCI state 3 | FCI state 4 |
|---|---|---|---|---|
| 0.000 | −15.799590 | −15.551083 | −15.551083 | −15.544641 |
| 0.250 | −15.794413 | −15.549756 | −15.542536 | −15.537486 |
| 0.500 | −15.784419 | −15.555851 | −15.535824 | −15.533966 |
| 0.750 | −15.771595 | −15.566635 | −15.540842 | −15.528096 |
| 1.000 | −15.757969 | −15.582025 | −15.552791 | −15.527076 |
| 1.250 | −15.744501 | −15.601268 | −15.570456 | −15.531302 |
| 1.500 | −15.730658 | −15.622205 | −15.591043 | −15.539058 |
| 1.750 | −15.714843 | −15.641867 | −15.610953 | −15.547214 |
| 2.000 | −15.695277 | −15.657547 | −15.626957 | −15.552573 |
| 2.250 | −15.671020 | −15.667490 | −15.636898 | −15.552936 |
| 2.500 | −15.670990 | −15.643186 | −15.639670 | −15.547310 |
| 2.750 | −15.668366 | −15.635006 | −15.623460 | −15.575085 |
| 2.825 | −15.666577 | −15.632203 | −15.626093 | −15.574018 |
| 3.000 | −15.661185 | −15.645845 | −15.623323 | −15.568085 |
| 3.250 | −15.680911 | −15.652848 | −15.605626 | −15.593049 |
| 3.500 | −15.714012 | −15.648450 | −15.615478 | −15.583506 |
| 3.750 | −15.741392 | −15.649889 | −15.633425 | −15.575981 |
| 4.000 | −15.757451 | −15.649935 | −15.641720 | −15.599588 |

TABLE VI: Symmetry mapping ($C_1 \to C_{2v}$) of the FCI states along the PES.

| region (values in $a_0$) | | | $\Gamma_1$ | $\Gamma_2$ | $\Gamma_3$ |
|---|---|---|---|---|---|
| 0 | $< x \leq$ | 0.25 | $^1A_1$ | $^3B_1$ | $^3A_2$ |
| 0.25 | $< x \leq$ | 2.25 | $^1A_1$ | $^3B_1$ | $^1B_1$ |
| 2.25 | $< x \leq$ | 2.5 | $^3B_1$ | $^1A_1$ | $^1B_1$ |
| 2.5 | $< x \leq$ | 2.825 | $^3B_1$ | $^1B_1$ | $^1A_1$ |
| 2.825 | $< x \leq$ | 3.0 | $^3B_1$ | $^1A_1$ | $^1B_1$ |
| 3.0 | $< x \leq$ | 3.25 | $^1A_1$ | $^3B_1$ | $^1B_1$ |
| 3.25 | $< x \leq$ | 4.0 | $^1A_1$ | $^3B_1$ | $^3B_2$ |

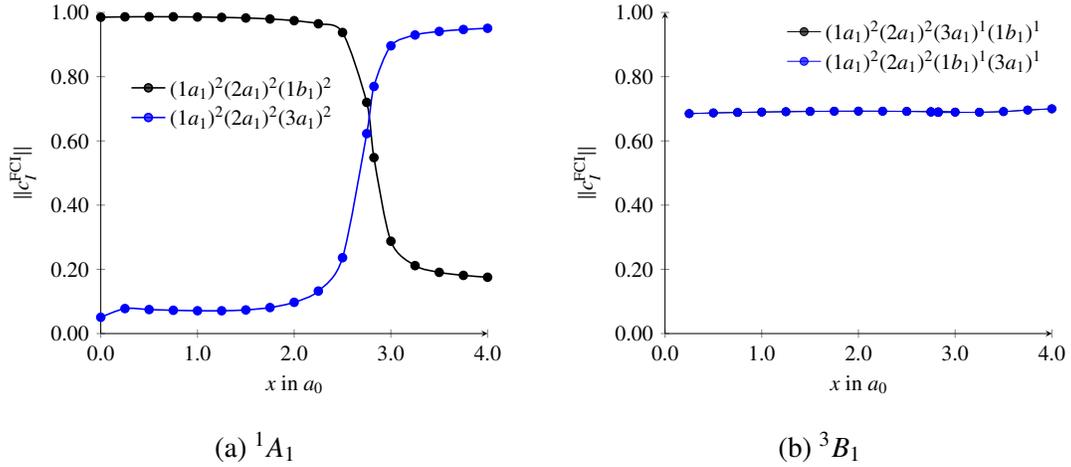

(a) $^1A_1$      (b) $^3B_1$

FIG. 1: Absolute FCI coefficients of the two dominant determinants based on CAS(2,2)SCF orbitals for the individual states.

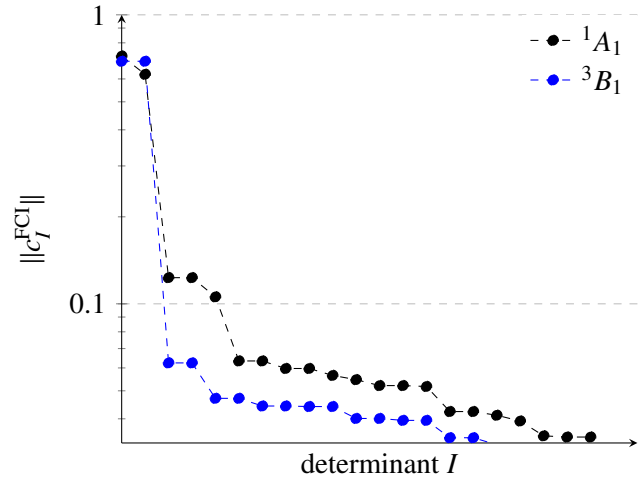

FIG. 2: Values of FCI coefficients at $x = 2.75\ a_0$ using CAS(2,2)SCF orbitals for the individual states.

## C. bivar-MR(E)CC results

The energy differences of different bivar-MR(E)CCSD method are listed in Tab. VII. The errors of the bivar-MRCCSD density operator relative to FCI results are depicted and compared to Hartree–Fock values in Fig. 3. It can be seen, that by increasing the dimension of the external space by including the first order interacting space, or by increasing the model space, the accuracy of the density operator is significantly improved. The bivar-MRECCSD(2,2)FOIS values are relatively similar to the bivar-MRCCSD(2,2)FOIS values, slightly inferior in the singlereference, and slightly superior in the multireference region, an observation which also has been described for the singlereference variants of these CC models.

Tab. VIII shows bivar-MRCCSD(2,2)FOIS energies based on CAS(2,2)SCF orbitals which were optimized for different electronic states. Apparently, one can get arbitrarily close to the FCI energy, and even below. The respective errors in the density operator are shown in Tab. IX, suggesting that the energies are not a good indicator for accuracy if the state of interest is an excited state (cf. values at $x = 2.75\ a_0$).

Finally, results on the $^3B_1$ states are shown in Tab. X. Concerning the total-spin expectation value, we note that the bivar-MRCCSD values are very close to the FCI value ($S(S + 1) = 2$).

TABLE VII: Energy differences of bivar-MR(E)CCSD computations with respect to FCI in $mE_\mathrm{H}$ for the $^1A_1$ state using the respective CASSCF orbitals for the $^1A_1$ state.

| $x$ in $a_0$ | bivar-MRCCSD(2,2) | bivar-MRCCSD(2,2)FOIS | bivar-MRECCSD(2,2)FOIS | bivar-MRCCSD(4,6) |
|---|---|---|---|---|
| 0.000 | 0.415 | 0.368 | 0.357 | 0.209 |
| 0.250 | 0.373 | 0.146 | 0.150 | 0.083 |
| 0.500 | 0.370 | 0.157 | 0.164 | 0.094 |
| 0.750 | 0.380 | 0.184 | 0.198 | 0.125 |
| 1.000 | 0.394 | 0.217 | 0.242 | 0.177 |
| 1.250 | 0.413 | 0.251 | 0.291 | 0.232 |
| 1.500 | 0.435 | 0.278 | 0.331 | 0.271 |
| 1.750 | 0.466 | 0.294 | 0.359 | 0.297 |
| 2.000 | 0.520 | 0.311 | 0.387 | 0.325 |
| 2.250 | 0.646 | 0.355 | 0.443 | 0.394 |
| 2.500 | 1.182 | 0.532 | 0.551 | 0.512 |
| 2.750 | 6.537 | 1.547 | 1.458 | 2.020 |
| 2.825 | 2.628 | 0.579 | 0.541 | 1.264 |
| 3.000 | 0.863 | 0.398 | 0.329 | 0.511 |
| 3.250 | 0.427 | 0.283 | 0.253 | 0.207 |
| 3.500 | 0.309 | 0.219 | 0.205 | 0.147 |
| 3.750 | 0.239 | 0.171 | 0.163 | 0.124 |
| 4.000 | 0.189 | 0.134 | 0.130 | 0.113 |

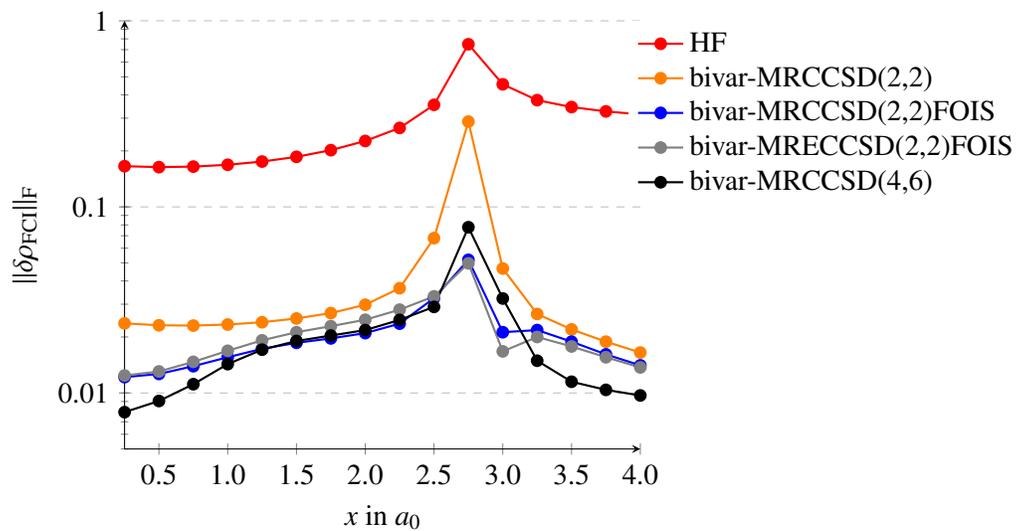

FIG. 3: Density operator error comparison ($^1A_1$ state, $^1A_1$ orbitals).

TABLE VIII: Energy differences $\Delta E_{\text{FCI}}$ in $mE_{\text{H}}$ of bivar-MRCCSD(2,2)fois computation for the $^1A_1$ state using CAS(2,2)SCF orbitals optimized for different states.

| $x$ in $a_0$ | $^1A_1$ | $^1A_1 + {}^3B_1$ (1:1) | $^3B_1$ | $^1B_1$ |
|---|---|---|---|---|
| 0.250 | 0.146 | 0.229 | 0.202 | 0.328 |
| 0.500 | 0.157 | 0.220 | 0.189 | 0.161 |
| 0.750 | 0.184 | 0.218 | 0.177 | 0.140 |
| 1.000 | 0.217 | 0.215 | 0.162 | 0.218 |
| 1.250 | 0.251 | 0.212 | 0.142 | 0.075 |
| 1.500 | 0.278 | 0.210 | 0.122 | 0.034 |
| 1.750 | 0.294 | 0.214 | 0.103 | −0.017 |
| 2.000 | 0.311 | 0.236 | 0.086 | −0.093 |
| 2.250 | 0.355 | 0.300 | 0.091 | −0.195 |
| 2.500 | 0.532 | 0.514 | 0.231 | −0.248 |
| 2.750 | 1.547 | 1.511 | 1.272 | 0.358 |
| 2.830 | 0.579 | 0.745 | 3.172 | 0.659 |
| 3.000 | 0.398 | 0.412 | 0.341 | 0.281 |
| 3.250 | 0.283 | 0.285 | 0.143 | 0.158 |
| 3.500 | 0.219 | 0.220 | 0.197 | 0.134 |
| 3.750 | 0.171 | 0.171 | 0.193 | 0.101 |
| 4.000 | 0.134 | 0.135 | 0.169 | 0.072 |

TABLE IX: Density operator error $\|\delta\rho_{\text{FCI}}\|_F$ comparison of bivar-MRCCSD(2,2)FOIS results based on CAS(2,2)SCF orbitals optimized for different electronic states.

| $x$ in $a_0$ | $^1A_1$ | $^1A_1 + {}^3B_1$ (1:1) | $^3B_1$ | $^1B_1$ |
|---|---|---|---|---|
| 0.250 | 0.012 | 0.017 | 0.016 | 0.023 |
| 0.500 | 0.013 | 0.017 | 0.015 | 0.014 |
| 0.750 | 0.014 | 0.016 | 0.015 | 0.013 |
| 1.000 | 0.016 | 0.016 | 0.014 | 0.016 |
| 1.250 | 0.017 | 0.016 | 0.013 | 0.010 |
| 1.500 | 0.019 | 0.016 | 0.012 | 0.007 |
| 1.750 | 0.020 | 0.017 | 0.011 | 0.001 |
| 2.000 | 0.021 | 0.018 | 0.011 | 0.009 |
| 2.250 | 0.024 | 0.022 | 0.012 | 0.016 |
| 2.500 | 0.032 | 0.032 | 0.016 | 0.037 |
| 2.750 | 0.052 | 0.033 | 0.108 | 0.206 |
| 3.000 | 0.021 | 0.026 | 0.033 | 0.047 |
| 3.250 | 0.022 | 0.023 | 0.025 | 0.014 |
| 3.500 | 0.019 | 0.020 | 0.023 | 0.021 |
| 3.750 | 0.016 | 0.017 | 0.020 | 0.023 |
| 4.000 | 0.014 | 0.014 | 0.017 | 0.026 |

TABLE X: Results of the bivar-MRCCSD computation on the $^3B_1$ state of BeH$_2$ using the respective CASSCF orbitals. Energy differences $\Delta E_{\text{FCI}}$ are given in $mE_\text{H}$ and total-spin expectation values $\langle S^2 \rangle$ in $\hbar^2$.

| | MRCCSD(2,2)FOIS | | MRCCSD(4,6) | |
|---|---|---|---|---|
| $x$ in $a_0$ | $\Delta E_{\text{FCI}}$ | $\langle S^2 \rangle$ | $\Delta E_{\text{FCI}}$ | $\langle S^2 \rangle$ |
| 0.250 | 2.631 | 1.931 | 2.029 | 1.965 |
| 0.500 | 3.282 | 1.968 | 1.901 | 1.987 |
| 0.750 | 3.500 | 1.978 | 1.767 | 1.992 |
| 1.000 | 3.610 | 1.982 | 1.866 | 1.993 |
| 1.250 | 3.692 | 1.983 | 1.954 | 1.992 |
| 1.500 | 3.752 | 1.982 | 1.998 | 1.992 |
| 1.750 | 3.785 | 1.979 | 2.055 | 1.991 |
| 2.000 | 3.798 | 1.981 | 2.099 | 1.991 |
| 2.250 | 3.815 | 1.980 | 2.068 | 1.992 |
| 2.500 | 3.868 | 1.981 | 1.971 | 1.993 |
| 2.750 | 3.997 | 1.985 | 1.846 | 1.995 |
| 2.825 | 4.056 | 1.986 | 1.827 | 1.995 |
| 3.000 | 4.229 | 1.986 | 1.751 | 1.996 |
| 3.250 | 4.402 | 1.993 | 1.635 | 1.998 |
| 3.500 | 3.847 | 1.996 | 1.518 | 1.999 |
| 3.750 | 2.575 | 1.998 | 1.543 | 1.999 |
| 4.000 | 1.584 | 1.999 | 1.228 | 1.999 |

## D. Orbital optimization

The influence of bivariational active-orbital optimization on the energy of the BeH$_2$ molecule at $x = 2.75\ a_0$ for the $^3B_1$ state is shown in Fig. 4. Similar to the results for $^1A_1$ discussed in the main text, the orbital optimization does not change the bivar-(OA)MRCCSD(4,6) energies significantly.

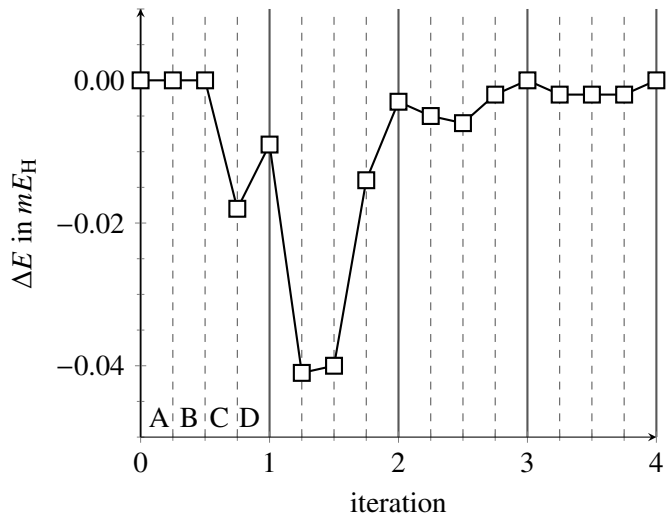

FIG. 4: Deviation of a bivar-OAMRCCSD(4,6) from a bivar-MRCCSD(4,6) computation for the BeH$_2$ model at $x = 2.75\ a_0$ in $mE_\mathrm{H}$ ($^3B_1$ state). Steps per iteration: A (amplitude equations), B (lambda equations), C (orbital optimization), D (eigenvalue problem).

### E. Properties

The accuracy of first-order properties computed with the bivar-MRCCSD method has been assessed by calculating the dipole moment of BeH$_2$ (for values of H–F, see Section II) and comparing to FCI values. Tab. XI lists these values and includes values of other quantum chemical methods for comparison. In particular in the multireference region, the bivar-MRCCSD results are superior compared to the singlereference CC and CAS(2,2)SCF values.

TABLE XI: Comparison of error norms $\|\Delta m_{\text{FCI}}\|_2$ for the electronic dipole moment of the BeH$_2$ model system computed with different methods.

| $x$ in $a_0$ | RHF   | CAS(2,2)SCF | RHF-CCSD | RHF-CCSDT | MRCCSD(4,6) | MRCCSD(2,2)FOIS |
|--------------|-------|-------------|----------|-----------|-------------|-----------------|
| 0.000        | 0.000 | 0.000       | 0.000    | 0.000     | 0.000       | 0.000           |
| 0.250        | 0.027 | 0.020       | 0.000    | 0.000     | 0.001       | 0.001           |
| 0.500        | 0.056 | 0.041       | 0.000    | 0.000     | 0.000       | 0.001           |
| 0.750        | 0.087 | 0.068       | 0.000    | 0.000     | 0.001       | 0.002           |
| 1.000        | 0.120 | 0.098       | 0.001    | 0.000     | 0.002       | 0.002           |
| 1.250        | 0.153 | 0.128       | 0.002    | 0.000     | 0.003       | 0.002           |
| 1.500        | 0.183 | 0.156       | 0.003    | 0.001     | 0.004       | 0.001           |
| 1.750        | 0.212 | 0.185       | 0.004    | 0.001     | 0.005       | 0.001           |
| 2.000        | 0.250 | 0.221       | 0.005    | 0.002     | 0.006       | 0.000           |
| 2.250        | 0.311 | 0.277       | 0.003    | 0.003     | 0.009       | 0.003           |
| 2.500        | 0.481 | 0.342       | 0.030    | 0.009     | 0.020       | 0.018           |
| 2.750        | 1.682 | 0.742       | 0.401    | 0.365     | 0.136       | 0.071           |
| 2.825        | 1.305 | 0.411       | 0.169    | 0.112     | 0.089       | 0.016           |
| 3.000        | 0.544 | 0.222       | 0.027    | 0.007     | 0.014       | 0.001           |
| 3.250        | 0.322 | 0.154       | 0.002    | 0.000     | 0.000       | 0.002           |
| 3.500        | 0.236 | 0.118       | 0.001    | 0.000     | 0.001       | 0.002           |
| 3.750        | 0.180 | 0.093       | 0.001    | 0.000     | 0.001       | 0.002           |
| 4.000        | 0.137 | 0.072       | 0.001    | 0.000     | 0.001       | 0.001           |



% \makeatletter
% \newcommand*{\balancecolsandclearpage}{%
%   \close@column@grid
%   \cleardoublepage
%   \twocolumngrid
% }
% \makeatother

% \balancecolsandclearpage
% \includepdf[pages=-,pagecommand={\balancecolsandclearpage\thispagestyle{empty}}]{si.pdf}

% %\includepdf[pages=-,pagecommand={\clearemptydoublepage\thispagestyle{empty}}]{si.pdf}

\end{document}